\documentclass[
reprint,superscriptaddress, amsmath,amssymb,aps,prl,footinbib
]{revtex4-2}

\usepackage{graphicx}
\usepackage{dcolumn}
\usepackage{mathtools} 
\usepackage{bm} 
\usepackage{bbold}
\usepackage{bbm}
\usepackage{tensor}
\usepackage{braket}
\usepackage{xcolor}
\usepackage{subcaption}
\usepackage{ragged2e}
\DeclareCaptionJustification{justified}{\justifying}
\usepackage[normalem]{ulem}
\usepackage[mathlines]{lineno}
\usepackage{hyperref} 
\hypersetup{colorlinks=true,linkcolor=blue,citecolor=blue,urlcolor=blue}
\usepackage[nameinlink]{cleveref}

\newcommand{\eff}{{\text{eff}}}

\newcommand{\labelphantom}[1]{
  \parbox{0pt}{\phantomsubcaption\label{#1}}
}

\allowdisplaybreaks

\begin{document}

\title{Entanglement generation in weakly-driven arrays of multilevel atoms via dipolar interactions}

\author{Sanaa Agarwal}
\email{sanaa.agarwal@colorado.edu}
\affiliation{
JILA, NIST, Department of Physics, University of Colorado, Boulder, CO 80309, USA}
\affiliation{Center for Theory of Quantum Matter, University of Colorado, Boulder, CO 80309, USA}

\author{A. Pi\~neiro Orioli}
\affiliation{QPerfect, 23 Rue du Loess, 67000 Strasbourg, France}
\affiliation{University of Strasbourg and CNRS, CESQ and ISIS (UMR 7006), 67000 Strasbourg, France}

\author{J. K. Thompson}
\affiliation{
JILA, NIST, Department of Physics, University of Colorado, Boulder, CO 80309, USA}

\author{A. M. Rey}
\email{arey@jilau1.colorado.edu}
\affiliation{
JILA, NIST, Department of Physics, University of Colorado, Boulder, CO 80309, USA}
\affiliation{Center for Theory of Quantum Matter, University of Colorado, Boulder, CO 80309, USA}

\date{\today}

\begin{abstract}
We investigate the driven-dissipative dynamics of 1D and 2D arrays of multilevel atoms interacting via dipole-dipole interactions and trapped at subwavelength scales. Here we show  that in the weakly driven low excitation regime,  multilevel atoms,
in contrast to two-level atoms, can become strongly entangled. The entanglement manifests as the growth of collective spin-waves in the ground state manifold, and survives even after turning off the drive. We propose to use the $\sim 2.9~\mu$m transition between $\rm ^3{\rm P}_2 \leftrightarrow \, ^3{\rm D}_3$ in $\rm  ^{88}Sr$ with $\rm 389~nm$ trapping light as an ideal experimental platform for validating our predictions and as a novel quantum interface for the exploration of complex many-body phenomena emerging from light-matter interactions.

\end{abstract}

\maketitle

\emph{Introduction}---Structured arrays of atoms trapped at subwavelength scales are emerging as unique  quantum platforms where strong photon-mediated dipolar interactions between atoms allow to significantly modify atomic lifetimes and their radiance properties \cite{dickeCoherenceSpontaneousRadiation1954,lehmbergRadiationAtomSystem1970,grossSuperradianceEssayTheory1982,weissSubradianceRadiationTrapping2018,Chang2018RevMod,ruiSubradiantOpticalMirror2020}. As a result, these systems have been predicted to have untapped applications as  quantum memories \cite{facchinettiStoringLightSubradiant2016,asenjo-garciaExponentialImprovementPhoton2017,jenPhaseimprintedMultiphotonSubradiant2017,ballantineSubradianceprotectedExcitationSpreading2020,santosGeneratingLonglivedEntangled2021}, quantum simulators of rich many-body physics \cite{olmosLongRangeInteractingManyBody2013,jonesModifiedDipoledipoleInteraction2018,needhamSubradianceprotectedExcitationTransport2019a,moreno-cardonerSubradianceenhancedExcitationTransfer2019}, to realize topological phases of matter \cite{Syzranov2014, syzranovEmergentWeylExcitations2016,perczelTopologicalQuantumOptics2017,bettlesTopologicalPropertiesDense2017,perczelPhotonicBandStructure2017,zhangTunableTopologicallyprotectedSuper2019,zhangTunableSuperSubradiant2019}, and as a way to improve atomic clocks \cite{changControllingDipoledipoleFrequency2004,henrietCriticalOpensystemDynamics2019,quSpinSqueezingManybody2019}. So far, most work investigating such phenomena has been restricted to atoms with a unique electronic ground state, especially two-level atoms, in the weak excitation limit \cite{guerinSubradianceLargeCloud2016,browaeys_subradiance_single_exc}.
In this limit, correlations are suppressed and a simple mean-field treatment, or classical dipoles approximation, is usually enough to capture the physics \cite{Bienaime2013a}.

\begin{figure}[ht!]
\labelphantom{fig:fig1a}
\labelphantom{fig:fig1b}
\labelphantom{fig:fig1c}
\labelphantom{fig:fig1d}
\includegraphics[width=1\linewidth]{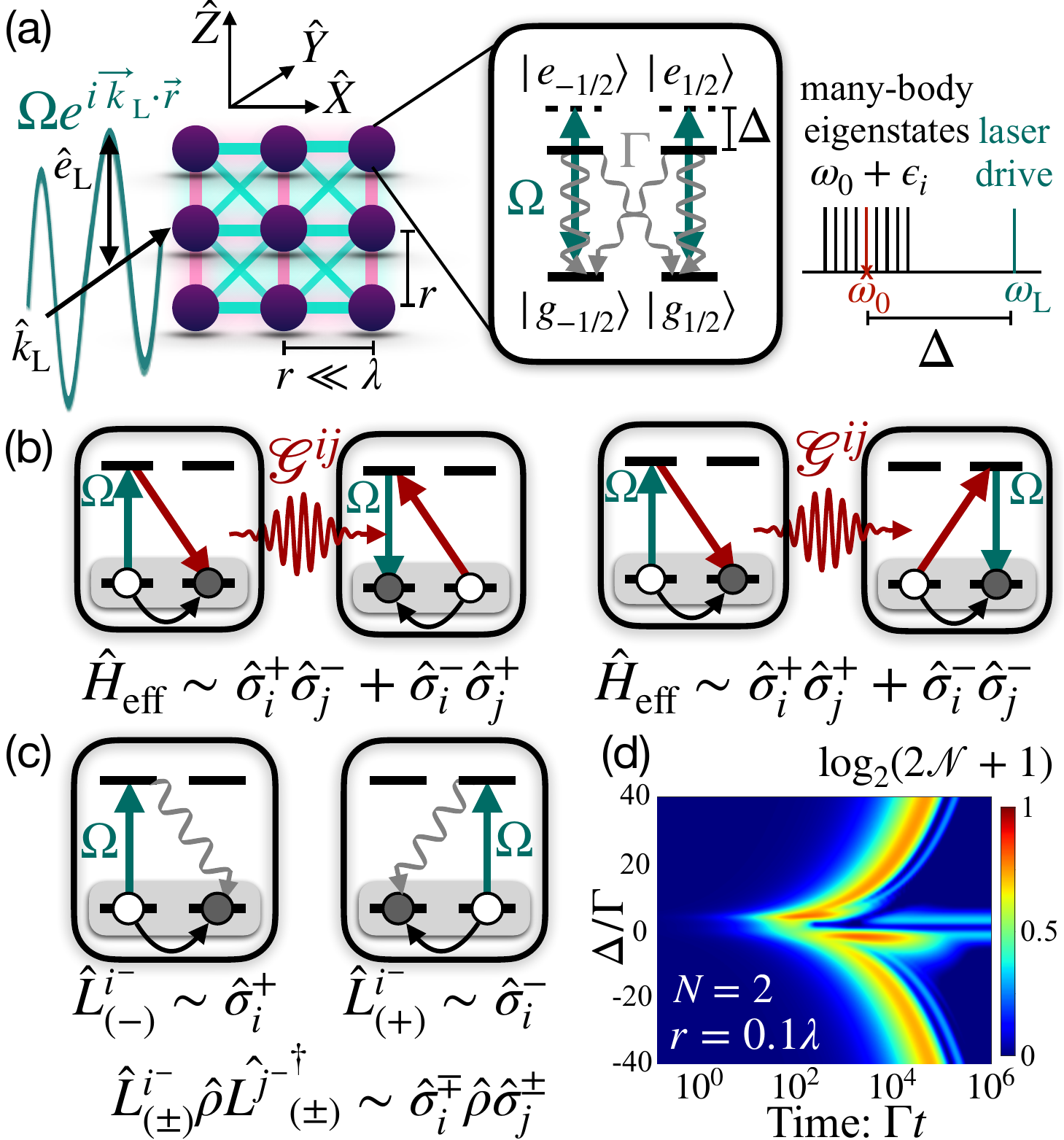}
\caption{ \textbf{Multilevel atomic array (2D) interacting via dipolar interactions}. (a) The subwavelength ($r\ll\lambda$) array is continuously driven weakly ($\Omega\ll\Gamma$) with detuning $\Delta$ from the single-atom transition with linewidth $\Gamma$. By adiabatically eliminating the excited states, the effective two-level ground state subspace processes capture the terms in (b) the effective Hamiltonian, $\hat H_{\eff}$, and (c) the effective jump operators, $\hat L^i_{(\pm)}$, up to lowest-order (see Eq.~(\ref{eq:Heff_large_det})). We show only the relevant terms here. (d) Entanglement ($\log_{2}(2\mathcal{N}+1)$) dynamics in the ground-state subspace, obtained from the effective model in Eq.~(\ref{eq:GSM_master_eq}). Blue (pink) links in (a) show negative (positive) interaction coefficients $\mathcal{C}_{ij}^y$ between neighboring atoms that determine the spin-spin correlations.} 
\end{figure}

In this paper, we study the dynamics of weakly-driven arrays of four-level atoms with two degenerate ground and excited states.
Multilevel atoms are quite complex even in the weak-excitation limit due to the exponential size of the ground state manifold and have, therefore, remained under-explored so far.
Nevertheless, recent theoretical work on multilevel systems in arrays
\cite{munroPopulationMixingDue2018a, asenjo-garciaOpticalWaveguidingAtomic2019a, pineiroorioliDarkStatesMultilevel2019, pineiroorioliSubradianceMultilevelFermionic2020, cidrimDipoledipoleFrequencyShifts2021, sutherlandSuperradianceInvertedMultilevel2017, sutherlandDegenerateZeemanGround2017, hebenstreitSubradianceEntanglementAtoms2017} and cavities \cite{reiterDrivingTwoAtoms2012, orioliEmergentDarkStates2021} has begun to uncover some of the rich many-body phenomena enabled by the ground state degeneracy that is absent in two-level atoms.

Our main finding is that, in the weak-excitation limit of driven-dissipative dynamics, in contrast to two-level atoms, four-level atoms can become strongly entangled at short inter-particle distances $r< \lambda/(2\pi)$, where $\lambda$ is the transition wavelength.
Such distances are required to enhance the coherent part of dipolar interactions relative to spontaneous emission, which destroys the correlations.
The generated entanglement lives in the ground state manifold and hence, can be simply stored by turning off the drive, making such states appealing for quantum information applications.
Such strong correlations render common approaches such as cumulant approximations, which work for two levels, inapplicable here.
Nevertheless, we are able to gain some insight into the entanglement generation process by deriving simplified effective models acting on the ground state manifold only.
The resulting effective Hamiltonian and jump operators both contain effective spin-spin interaction terms stemming from multi-photon processes that lead to strong quantum correlations. By considering the large-detuning limit, we are able to truncate the interactions to up to two-body terms and obtain an anisotropic XY model. Unlike the large-detuning limit of the original two-level atoms, in which the interactions (and hence, the development of correlations) are suppressed compared to spontaneous emission,  we see the growth of collective spin-waves and spin-squeezing here. 
Finally, we propose an experimental setting to engineer the desired subwavelength arrays.
We propose to use the cycling $5s5p\, ^3 {\rm P}_2 \leftrightarrow 5s4d\,^3{\rm D}_3$  transition  of  $^{88}$Sr  atoms  at  $2.9~\mu$m with a natural line width of $\rm 57~kHz$ \cite{MassonPRXQ2024,SansonettiJPCRD2010,Hashiguchi2019,Bowden2019} and to trap the atoms using an optical lattice or tweezer setup with light at $\rm 389~nm$. 
Using standard AC Stark shifts and taking the several hundred seconds long lifetime \cite{yasudaLifetimeMeasurementMetastable2004} of $^3 {\rm P}_2$ into account, one can isolate a simple four-level system with $r/\lambda = 0.067$, where our predictions can be tested.

\emph{Multilevel dipolar master equation}---We consider an array of $N$ point-like multilevel atoms, pinned at their positions by either a deep optical lattice or a tweezer setup (Fig.~\ref{fig:fig1a}). 
The atoms have an optical transition of wavelength $\lambda\ (k_0=2\pi/\lambda)$, frequency $\omega_0$, and natural linewidth $\Gamma$, between a ground and an excited set of levels
with total angular momenta $F_g$ and $F_e$, respectively.
We label the Zeeman sublevels by
$\ket{g_{m_g}}$ and $\ket{e_{m_e}}$,
where $m_{g/e}=-F_{g/e},\dots,F_{g/e}$ is the magnetic number. 
We will focus on the four-level case with $F_g=F_e=1/2$ depicted in Fig.~\ref{fig:fig1a}; however, some of our findings are expected to generalize to other multilevel structures.

The array is weakly driven by a laser which is detuned from the atomic transition by a frequency $\Delta=\omega_{\rm L}-\omega_0$, has a Rabi frequency $\Omega$ ($\Omega \ll \Gamma $), wavevector  $\vec{k}$, and polarisation $\hat{e}_{\rm L}$. 
The effect of this drive is described by $(\hbar=1)$ 
\begin{equation}
\hat{H}_0 = - \Delta\sum_{i,m} \hat\sigma_{e_me_m}^{i} - \sum_{i,q} \big[\Omega\, (\hat{e}_{\rm L} \cdot \hat{e}_q^*)\,  {\hat {\mathcal D}}^{i^+}_{q} e^{i\vec{k} \cdot \vec{r}_i} + h.c. \big],
\end{equation}
where $\hat\sigma^j_{a_mb_n} \equiv |{a_m}\rangle_j {}_j\langle{b_n}|$, $a,b\in\{g,e\}$, and the sums run over all atoms $i$ and polarizations $q=0,\pm1$.
The laser only addresses the atomic transitions whose dipole moments have an overlap with the laser polarization through the term $(\hat{e}_{\rm L} \cdot \hat{e}_q^*)$, with $\hat{e}_0=\hat{Z}$ and $\hat{e}_{\pm1}=\mp(\hat{X}\pm i\hat{Y})/\sqrt{2}$. 
For simplicity, we drive the atoms with the laser polarisation pointing along the quantization axis and the laser wavevector perpendicular to the array (Fig.~\ref{fig:fig1a}) such that the driving Hamiltonian is $ \hat{H}_0 = - \Delta\sum_{i,m} \hat\sigma_{e_me_m}^{i} - \Omega\sum_{i} \big[{\hat {\mathcal D}}^{i^+}_{0} + h.c. \big] $.

The operator $ {\hat {\mathcal D}}^{i^+}_{q}$ ($ {\hat {\mathcal D}}^{i^-}_{q}=( {\hat {\mathcal D}}^{i^+}_{q})^\dagger$) is a generalized multilevel raising (lowering) operator defined as $ {\hat {\mathcal D}}^{i^+}_{q} = \sum_{n} C_n^q\, \hat\sigma_{e_{n+q} g_n}^i$, analogous to the two-level Pauli spin operators.
It describes the  change  in internal levels caused by the absorption/emission of a photon with polarization $q$. It  is a superposition of all transitions with $q=m_e-m_g$, weighted by the Clebsch-Gordan (CG) coefficient of the transition, $C_n^q=\langle F_g, n; 1,q | F_e,n+q \rangle$.

When the atoms become excited they interact via standard photon-mediated dipolar interactions given by
\begin{align}
\label{eq:Hdd}
\hat{H}_\text{d-d} =& - \sum_{qq',i\neq j} \Delta_{q,q'}^{ij}  {\hat {\mathcal D}}^{i^+}_{q} {\hat {\mathcal D}}^{j^-}_{q'},
\\
\label{eq:Ldd}
\mathcal{L}_\text{d-d}(\hat{\rho}) =& \sum_{qq',ij} \Gamma_{q,q'}^{ij} \big( 2  {\hat {\mathcal D}}^{j^-}_{q'} \hat\rho {\hat {\mathcal D}}^{i^+}_{q} -  \{ {\hat {\mathcal D}}^{i^+}_{q} {\hat {\mathcal D}}^{j^-}_{q'},\hat\rho\} \big),
\end{align}
where $\Delta_{q,q'}^{ij} = \hat{e}_q^{*^{\rm T}}\cdot {\rm Re}\mathit{G}(\vec{r}_{ij})\cdot\hat{e}_{q'}$,
$\Gamma_{q,q'}^{ij} = \hat{e}_q^{*^{\rm T}}\cdot{\rm Im}\mathit{G}(\vec{r}_{ij})\cdot\hat{e}_{q'}$, $\vec{r}_{ij}=\vec{r}_i-\vec{r}_j$,
and the photon exchange is mediated by the free-space electromagnetic Green's tensor, $\mathit{G}(\vec{r}) = \frac{3\Gamma}{4}\left[(\mathbbm{1}-\hat{r} \otimes \hat{r})\frac{e^{ik_0 r}}{k_0 r} + (\mathbbm{1}-3\hat{r} \otimes \hat{r})\left(\frac{i e^{ik_0 r}}{(k_0 r)^2} - \frac{ e^{ik_0 r}}{(k_0 r)^3}\right) \right]$, 
where $r\equiv|\vec{r}|$, and $\otimes$ denotes the outer product \cite{gross1982superradiance,jamesFrequencyShiftsSpontaneous1993a,lehmbergRadiationAtomSystem1970}.
The full dynamics of our system can thus be described by a master equation $\dot{\hat{\rho}}(t)  = -i [\hat{H}_0+\hat{H}_\text{d-d},\hat{\rho}(t)] + \mathcal{L}_\text{d-d}(\hat{\rho}(t))$ \cite{changControllingDipoledipoleFrequency2004,pineiroorioliSubradianceMultilevelFermionic2020}, where $\hat\rho$ is the atomic density matrix.

The Hamiltonian part, Eq.~(\ref{eq:Hdd}), describes a flip-flop interaction between two atoms with ${\hat {\mathcal D}}^{i^\pm}_{q}$ in place of the usual two-level $\hat\sigma_i^\pm$.
Note that atoms can exchange photons between transitions of orthogonal dipole moments, i.e., $\Delta_{q\neq q'}^{ij} \neq 0$.
The Lindbladian part, Eq.~(\ref{eq:Ldd}), describes correlated emission for $i\neq j$ and single-particle spontaneous emission for $i=j$.
Importantly, at short enough distances, $k_0 r \ll 1$, the coherent part scales as $\Delta_{q,q'}^{ij} \sim 1/{(k_0r)}^3$ and dominates over the incoherent part, which becomes constant, $\Gamma_{q,q'}^{ij}\rightarrow \Gamma_{q,q'}^{ii} = \Gamma/2\, \delta_{q,q'}$. 
Thus, we consider short subwavelength lattice spacings, $r<\lambda/(2\pi)$, in order to boost coherent dipolar interactions, which we find are the key to generating entanglement in the ground state manifold at weak driving.

\begin{figure}[t!]
\labelphantom{fig:fig2a}
\labelphantom{fig:fig2b}
\labelphantom{fig:fig2c}
\labelphantom{fig:fig2d}
\labelphantom{fig:fig2e}
\labelphantom{fig:fig2f}
\includegraphics[width=1\linewidth]{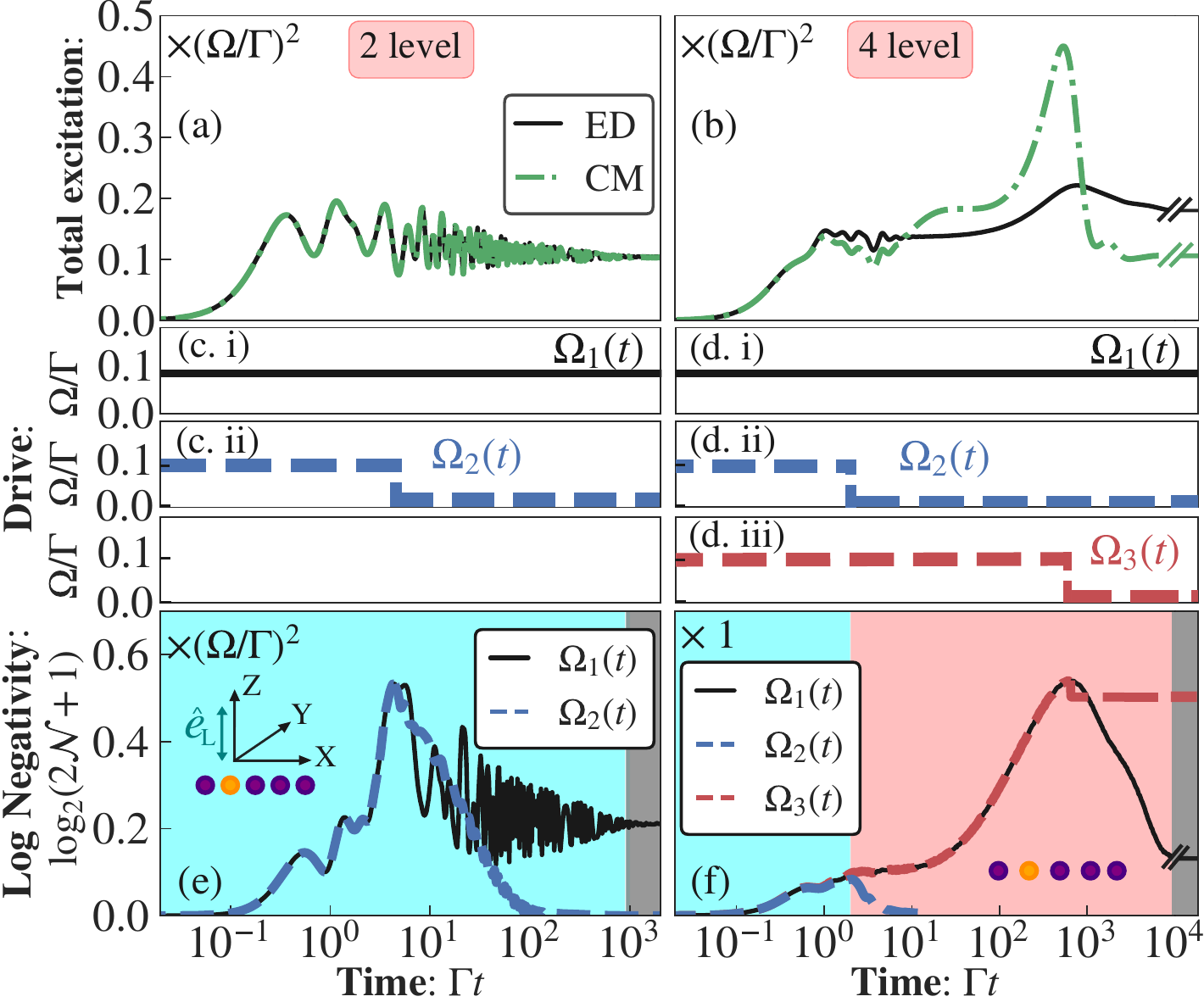}
\caption{(a, b): Dynamics of the total excited population calculated using exact diagonalization (ED, solid black) and cumulant method (CM, dash-dotted green) for a 1D array of $N=5$ atoms with $\Omega=0.1\Gamma,r=0.1\lambda,\Delta=-3\Gamma$. ED and CM agree for (a) two-level but not for (b) four-level atoms. (c.i-ii, d.i-iii) Driving protocols $\Omega_{1,2,3}(t)$, which correspond to turning off the drive in the different regimes shaded as grey, blue, and red in the panel below. (e, f) ED dynamics of bipartite log negativity between yellow and purple atoms in the array for (e) two-level and (f) four-level arrays. Note the $y$-axis in (f) is not scaled by $\Omega/\Gamma$. The axis breaks show a stop in the numerics before fully reaching the steady state (due to long runtimes).} \label{fig:N4_negativity_total_exc_wrt_time}
\end{figure}

\emph{Entanglement generation in multilevel vs two-level}---
In the weak drive regime ($\Omega\ll\Gamma$), the population of the excited state is suppressed by $(\Omega/\Gamma)^2$ and hence, the dynamics mostly takes place in the ground and single-excitation manifold. In this limit, the effective Hilbert space of two-level atoms scales linearly with $N$ and the system can be described as $N$ classical dipoles.
However, in the presence of multiple ground states, the size of the ground manifold grows exponentially with system size and the system can become highly entangled.

In Fig.~\ref{fig:N4_negativity_total_exc_wrt_time}, we use the logarithmic negativity $\log_2(2\mathcal{N}+1)$, which is a common bipartite entanglement measure for mixed states \cite{LewensteinPRA1998,VidalPRA2002,PlenioPRL2005,eisert2006entanglement} to quantify entanglement build-up.
It can be obtained by taking a partial transpose $\rho^{{\rm T}_A}$ over a subsystem $A$ and summing over its negative eigenvalues $\{\lambda_n\}$ as $\mathcal{N} = \sum_n (|\lambda_n|-\lambda_n)/2$.
Specifically, we compute the entanglement between one atom (typically a central atom) and the rest of the array, such that $\log_2(2\mathcal{N}+1)$ ranges from 0 (un-entangled) to 1.

For comparison purposes,  we consider $N=5$ two-level ($F_g=F_e=0$) and four-level ($F_g=F_e=1/2$) atoms, with a spacing $r=0.1\lambda$, subject to a weak drive, $\Omega=0.1\Gamma$, in a 1D configuration along the $\hat{X}$-axis (see Fig.~\ref{fig:fig2e}).
The atoms are initialized in a product state in the ground subspace, $\ket{g}^{\otimes N}$ (two-level) or $\left[(\ket{g_{-1/2}}+\ket{g_{1/2}})/\sqrt{2}\right]^{\otimes N}$ (four-level). Then, the drive is turned on and they are allowed to evolve until they reach the steady state. Moreover, we consider three different pulse shapes denoted as $\Omega_{1,2,3}(t)$, in which we turn off the drive at different times during the dynamics, as shown in Figs.~\ref{fig:fig2c}.i-ii, ~\ref{fig:fig2d}.i-iii.
We compare exact diagonalization (ED, solid black) results with a second-order cumulant expansion method (CM, dash-dotted green).

In the two-level system, the excited state population (Fig.~\ref{fig:fig2a}) and entanglement (Fig.~\ref{fig:fig2e}) remain small at all times and of order $(\Omega/\Gamma)^2\sim 10^{-2}$, in accordance with the weak-drive assumption. This entanglement is generated by correlations between ground-excited coherences, which are suppressed by $\Omega/\Gamma$.
Because of this, the entanglement vanishes as $\Omega/\Gamma\rightarrow0$, where mean-field becomes accurate \cite{CourteilleEPJD2010,BienameJMO2011}.
At small but finite $\Omega/\Gamma$, the small correlations present can be captured by the cumulant expansion, as demonstrated by its agreement with ED in Fig.~\ref{fig:fig2a}.

Similar to the two-level case, the excited state population of the four-level system is also suppressed by $(\Omega/\Gamma)^2$, as shown in Fig.~\ref{fig:fig2b}. 
However, the four-level array reaches a large entanglement that is not suppressed by the drive strength, as shown in Fig.~\ref{fig:fig2f} (note that the vertical axis is not scaled by $\Omega/\Gamma$). As a result, not only mean-field fails  in the $\Omega/\Gamma\rightarrow0$ limit but also CM (Fig.~\ref{fig:fig2b}). Since the latter is able  to capture two-body correlations, its failure signals the emergence of genuine multi-body correlations, which complicate the numerical simulation of these systems for large $N$.

We find that the entanglement dynamics of the four-level system can be divided into three qualitatively different time regimes marked in Fig.~\ref{fig:fig2f} by the shaded areas: (I, blue) build-up of excited state entanglement, (II, red) growth of ground state entanglement, and (III, gray) steady-state entanglement.

At short times of order $t_\text{I} \sim \Omega^{-1}$ (regime I, blue shaded), the drive generates coherences between the ground and excited states. This allows the atoms to interact and hence, build up entanglement between the dipoles. Since this entanglement depends on the ground-excited coherences, it vanishes as $\Omega/\Gamma \rightarrow 0$. To test if the correlations involve the excited state manifold, we perform another pulse shape, $\Omega_2(t)$ (Fig.~\ref{fig:fig2d}.ii), by turning off the drive in the middle of regime I. As expected, we find that spontaneous emission destroys the excited-ground correlations, disrupting the entanglement, as shown in Fig.~\ref{fig:fig2f} (dashed blue). 
This type of entanglement is analogous to the two-level case, as shown in Fig.~\ref{fig:fig2e} for the pulse shape in Fig.~\ref{fig:fig2c}.ii, and is well-described by CM as correlations are weak.

At intermediate times (regime II, red shaded) the four-level system develops entanglement in the ground subspace. The timescale of this regime is complicated to describe on-resonance, but in the large-detuning limit $(\Delta \gg \Gamma/(k_0 r)^3)$, it can be estimated up to the lowest order by treating the interactions perturbatively as $t_\text{II} \sim 3\Delta^2 (k_0 r)^3/(\Omega^2\Gamma)$.
As opposed to regime I, the entanglement in regime II remains finite as $\Omega/\Gamma\rightarrow0$ (see next section), and is preserved even after the drive is turned off. We observe this in Fig.~\ref{fig:fig2f} (red-dashed) by performing another pulse shape, $\Omega_3(t)$ (Fig.~\ref{fig:fig2d}.iii), in which the drive is turned off in the middle of regime II. This is because the excited state population is negligible and hence the ground state density matrix barely changes after the switch-off.

At very long times, $t\rightarrow\infty$, (regime III, gray shaded), higher order multi-photon processes become relevant while the drive remains on, until the four-level array  relaxes to a more complex mixed steady state at time $t\gg \Delta^2(k_0 r)^3/(\Omega^2\Gamma)$, as shown in Fig.~\ref{fig:fig2f}. 
Thus, turning off the drive at an earlier time preserves the entanglement, which may be lost in reaching the steady state.

The picture presented above qualitatively describes other parameter regimes too. In particular, Fig.~\ref{fig:fig1d} shows that the generation of large ground-state entanglement at intermediate times is a generic feature of the system for a wide range of detunings $\Delta$. For other cases, we refer to the SM~\cite{suppl_material}.

\emph{Effective ground-state model}---
To understand the ground state dynamics, we derive an effective model for just the ground-state subspace by treating the drive Hamiltonian, $\hat{V}_+ = -\sum_{i,q} \Omega_q^i {\hat {\mathcal D}}^{i^+}_{q} $, as a  perturbation that induces a weak coupling between the ground and excited sectors described by  the non-Hermitian Hamiltonian, $\hat H_{\rm NH} = -\Delta \sum_{i,m} \hat\sigma_{e_me_m}^i - \sum_{i,j,q,q'} \left( \Delta_{q,q'}^{ij} + i \Gamma_{q,q'}^{ij} \right){\hat {\mathcal D}}^{i^+}_{q}{\hat {\mathcal D}}^{j^-}_{q'}$. To second order in  $(\Omega/\Gamma)^2$ we obtain~\cite{reiterEffectiveOperatorFormalism2012a}
\begin{align}\label{eq:GSM_master_eq} 
\dot{\hat\rho} = -i[\hat{H}_{\eff}, \hat\rho] + \sum_{i,j,q,q'} \Gamma_{q,q'}^{ij}  \left[ 2{\hat L}^{j^-}_{q'}\hat\rho {\hat L}^{i^+}_{q}  -\{{\hat L}^{i^+}_{q} {\hat L}^{j^-}_{q'},\hat\rho\}\right],
\end{align}
where $\hat{H}_{\eff}= -\frac{1}{2} {\hat V}_- \left[ {\hat H}_{\rm NH}^{-1} + \left({\hat H}_{\rm NH}^{-1}\right)^\dagger \right]{\hat  V}_+,$ is an effective hermitian Hamiltonian (${\hat V}_-={\hat V}_+^\dagger $) and   ${\hat L}^{i^-}_{q} = {\hat {\mathcal D}}^{i^-}_{q} \hat{H}_{\rm NH}^{-1} \hat{V}_+$, effective jump operators 
both acting on the ground-state manifold (see SM~\cite{suppl_material} for details). The master equation has an overall scaling of $\sim\Omega^2/\Gamma$; thus, reducing $\Omega/\Gamma$ further only re-scales the timescale without affecting the generation of entanglement, unlike the original two-level case. 

In general, $\hat{H}_{\eff}$ and ${\hat L}^{i^-}_{q}$ are complicated quantities that may contain all possible $n$-body terms with $n\leq N$, originating from $(n-1)$-photon exchange processes. However, the physics simplifies in the large detuning limit $(\Delta\gg \Gamma/(k_0r)^3 \gg \Omega)$, where the role of $n>2$ terms is suppressed and the physics is dominated by up to two-body terms in the Hamiltonian and up to single-body terms in the jump operators as
\begin{equation}\label{eq:Heff_large_det}
\hat{ H}_{\rm eff} =  \sum_{i\neq j,\alpha} \mathcal{C}_{ij}^\alpha {\hat \sigma}^\alpha_i \hat{\sigma}^\alpha_j, \,\,
{\hat L}^{j^-}_{0} = \frac{\Omega}{3\Delta} \mathbbm{1}, \,\, {\hat L}^{j^-}_{\pm 1} = -\frac{\sqrt{2}\Omega}{3\Delta} {\hat \sigma}_\mp^j 
\end{equation}
where $\hat{\sigma}^{\alpha}_{i}$, $\alpha\in \{x,y\}$, are Pauli operators acting on the two-level ground subspace. The interaction coefficients are dominated by the near-field terms and given by  $ \mathcal{C}_{ij}^x = \frac{\Gamma \Omega^2}{12\Delta^2} \frac{\cos (k_0 |{\bf r}_{ij}|)}{\big(k_0 |{\bf r}_{ij}|\big )^3}$, $\mathcal{C}_{ij}^y = \mathcal{C}_{ij}^x\left(1 - 6({\hat r}_{ij}\cdot\hat{e}_+)^2\right)$. As shown in Fig.~\ref{fig:fig1b}, 
the lowest-order interaction terms in the Hamiltonian $\hat{H}_\text{eff}$ originate from a process where an atom $i$ absorbs a laser photon ($\Omega$), then exchanges the excitation via a vacuum-mode photon ($G^{ij}$) with another atom $j$, and atom $j$ finally emits a photon into the drive ($\Omega$). The final ground states of atoms $i$ and $j$ depend on their initial states and the atomic transitions involved in the photon exchange (red arrows). Up to lowest-order, the effective Lindblad terms (shown in Fig.~\ref{fig:fig1c}) can be described by processes where atoms get excited by the laser ($\Omega$) and subsequently decay to the ground state ($\Gamma$). The decay of an atom is affected by the presence of other atoms in the array, similar to the correlated emission from the excited state manifold.

Next, we consider the truncated Hamiltonian in Eq.~(\ref{eq:Heff_large_det}) to study the entanglement growth for large $N$. However, since the logarithmic negativity used above is difficult to experimentally access for large $N$, we shift our focus to spin-squeezing, which is relatively easy to measure experimentally and remains valid as an entanglement witness for open systems \cite{KitagawaPRA1993,winelandPRA1994,TothPRA2009}.

\begin{figure}[t!]
\includegraphics[width=\linewidth]{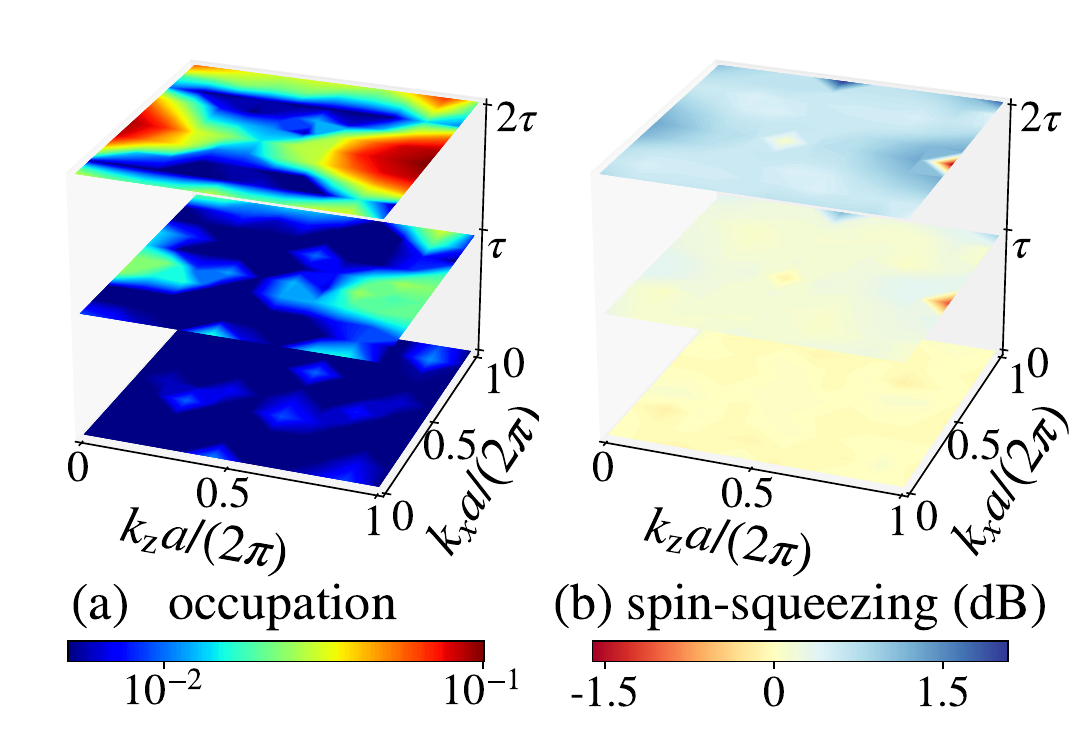}
\labelphantom{fig:fig3a}
\labelphantom{fig:fig3b}
\caption{\textbf{Spin-wave dynamics}: (a) Mode occupation $\langle n_{\vec{k}}(t)\rangle = \big(\langle \hat{\tilde{S}}_z^2 (\vec{k}) \rangle + \langle \hat{\tilde{S}}_y^2 (\vec{k})\rangle \big)/N - 1/2 $, and (b) spin-squeezing, $N\langle \hat{\tilde{S}}_{\phi_*}^2 (\vec{k})\rangle/\langle \hat{S}_x\rangle^2 $, of mode $\vec{k}$ for $\Omega\ll\Delta,r=0.1\lambda$, where $\hat{\tilde{S}}_{\phi}(\vec{k})=-\hat{\tilde{S}}_{z}(\vec{k}) \cos\phi + \hat{\tilde{S}}_{y}(\vec{k}) \sin\phi$ and optimal squeezing occurs at $\phi_*$. Data obtained using DTWA dynamics of the anisotropic XY model (Eq.~(\ref{eq:Heff_large_det})) for a $10\times 10$  array with $\hat{e}_{\rm L}=\hat{Z}$. The $z$-axis is time in units of $\tau=0.04\Delta^2/(\Gamma\Omega^2)$.} 
\end{figure}

\emph{Spin-spin correlations}---In the large detuning limit with $k_0 |\Vec{r}_{ij}|\ll 1$, the short-time dynamics is dominated by $\hat{H}_{\eff}$ and can be described via unitary dynamics. Dissipation kicks in later, reducing the amount of entanglement without changing the physics qualitatively (see SM \cite{suppl_material}).    $H_{\rm eff}$ describes an anisotropic XY model, which can be further simplified to an isotropic XY model $(\mathcal{C}^x_{ij}=\mathcal{C}_{ij}^y)$ by choosing a laser polarization along the 1D array $({\hat r}_{ij}\cdot\hat{e}_+)=0$. While this is not possible in 2D, the structure of spin-spin correlations is still controllable up to an extent by tuning the direction of the laser polarization in the $X-Z$ plane (see SM \cite{suppl_material}).  

We observe a non-trivial build-up  of  spin-spin correlations in both 1D and 2D, which can be described analytically at early times using Spin-Wave Analysis (SWA) \cite{HP1940,RoscildePRB2023} (see SM \cite{suppl_material} for derivation). 
We define the spin-wave operators as the Fourier Transform (FT) of the position-space operators as $\hat{\tilde{S}}_{z} (\vec{k})=-\sum_i (\hat \Sigma_i^{+} e^{i\vec{k}\cdot \vec{r}_i} + \hat\Sigma_i^{-} e^{-i\vec{k}\cdot \vec{r}_i})/2$ and $\hat{\tilde{S}}_{y} (\vec{k})=-i\sum_i (\hat\Sigma_i^{+} e^{i\vec{k}\cdot \vec{r}_i} -\hat\Sigma_i^{-} e^{-i\vec{k}\cdot \vec{r}_i})/2$, with the initial magnetisation along $\hat{{S}}_{x} = \sum_i \hat\sigma_i^{x}/2  $, where $\hat \Sigma_i^{+}=(-\hat\sigma_i^z + i\hat\sigma_i^y)/2$ is the raising operator along $\hat{x}$ in the Bloch sphere. 
SWA allows us to describe linear excitations above the collective initial condition and capture the generation of correlations in momentum-space via the spin-structure factor, $\langle n_{\vec{k}}(t)\rangle = \big(\langle \hat{\tilde{S}}_z^2 (\vec{k}) \rangle + \langle \hat{\tilde{S}}_y^2 (\vec{k})\rangle \big)/N - 1/2 $.

The 1D case is dominated by the nearest-neighbor interactions, which are short-range, and due to their sign, the eigenfrequency of excitations is minimum at mode $k_x=\pi/a$ ($a=$ lattice constant) and is in fact negative, making this mode unstable (see SM \cite{suppl_material}).  
In 2D, this mode gets stabilized by next-to-next-nearest neighbor terms due to changing signs of the coefficients (Fig.~\ref{fig:fig1a}), but it still has the lowest frequency, and thus, shows the largest growth in the structure factor 
(Fig.~\ref{fig:fig3a}). 
In Figs.~\ref{fig:fig3a},~\ref{fig:fig3b}, we show the unitary dynamics of a $10\times 10$ array using the Discrete Truncated Wigner Approximation (DTWA) \cite{SchachenmayerPRX2015,Zhu_2019}. We validate DTWA for this system with ED and SWA (see SM \cite{suppl_material}).
We define a spin-wave operator at angle $\phi$ in phase-space as $\hat{\tilde{S}}_{\phi}(\vec{k})=-\hat{\tilde{S}}_{z}(\vec{k}) \cos\phi + \hat{\tilde{S}}_{y}(\vec{k}) \sin\phi$. The spin-squeezing parameter for a mode-$\vec{k}$ is defined as \cite{winelandPRA1994,rosario2024detecting} $N\langle \hat{\tilde{S}}_{\phi_*}^2 (\vec{k})\rangle/\langle \hat{S}_x\rangle^2 $ ($\langle \hat{\tilde{S}}_{\phi} (\vec{k}) \rangle = 0$ for all $\phi$), where optimal squeezing occurs at $\phi_*$ (see SM \cite{suppl_material} for details). In Fig.~\ref{fig:fig3b}, we see the spin-squeezing growing for the mode with the largest occupation. 
The anti-squeezing, $N\langle \hat{\tilde{S}}_{\phi_*+\pi/2}^2 (\vec{k})\rangle/\langle \hat{S}_x\rangle^2 $, is along  the orthogonal spin quadrature and shows the opposite behavior. The non-trivial spin-structure factor  in our system is similar to recent observations in interacting dipolar arrays \cite{dominguezarxiv2023,BilitewskiPRA2023}, nevertheless in a more complex setting here.

\begin{figure}[t!]
\labelphantom{fig:fig4a}
\labelphantom{fig:fig4b}
\begin{center}
\includegraphics[width=0.48\linewidth]{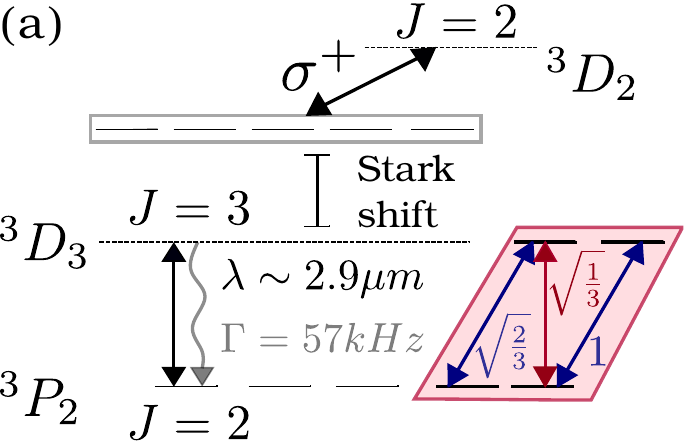}
\includegraphics[width=0.48\linewidth]{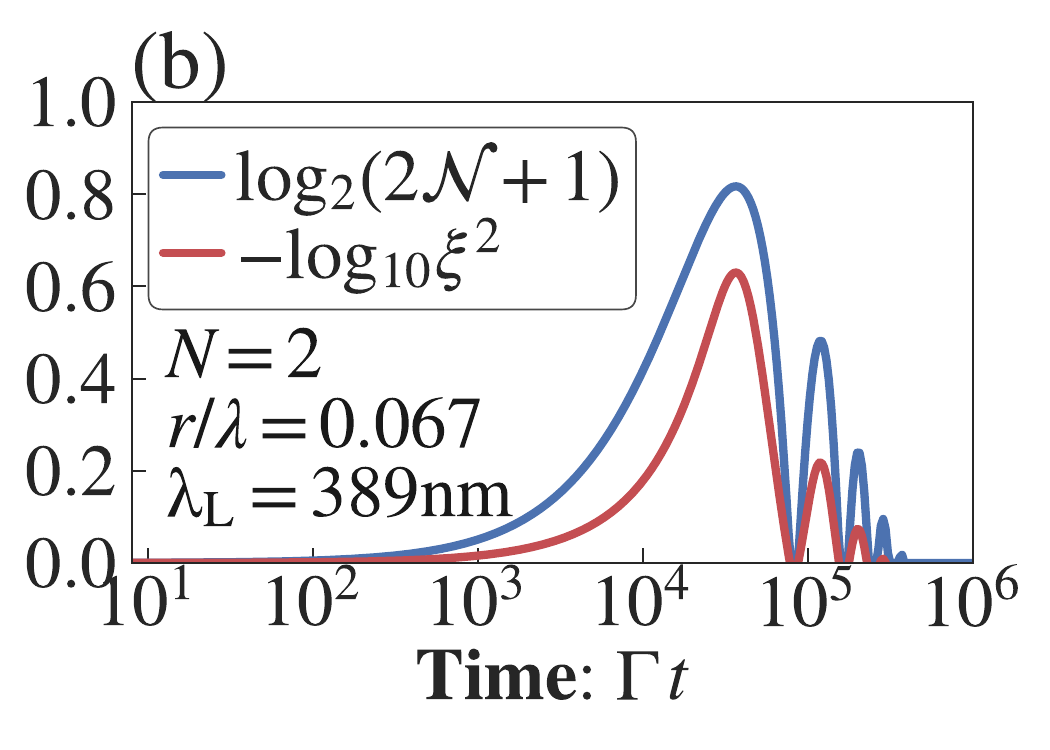}
\end{center}
\caption{\textbf{Experimental proposal using $\rm ^3{\rm P}_2 \leftrightarrow \,^3{\rm D}_3$ transition in $\rm ^{88}Sr$ atoms:} (a) Applying an AC Stark shift ($\sigma^+$) to levels $m_J=\{-3,-2,-1,0,1\}$ of $\rm ^3{\rm D}_3$ manifold to obtain an effective four-level atom (shaded area). (b) Log negativity and spin-squeezing \cite{TothPRA2009} for a system of two $\rm ^{88}$Sr atoms with $\Omega=0.1\Gamma, \Delta=27\Gamma$.} \label{fig:Sr_proposal}
\end{figure}

\emph{Experimental realization}---We propose to test our predictions
using an optical tweezer/lattice experimental setup with $^{88}$Sr atoms. $^{88}$Sr has a cyclic transition between $5s5p\, ^3 {\rm P}_2 \leftrightarrow 5s4d\,^3{\rm D}_3$ with wavelength $\rm \lambda \sim 2.9~\mu m$ and linewidth $\rm 57~kHz$ \cite{Hashiguchi2019}. 
The advantage of this transition is that the large $\lambda$ enables us to prepare very short inter-atomic distances using an optical lattice
with a laser of wavelength $\rm \lambda_L=389~nm$, i.e.~$r=\lambda_{\rm L}/2=0.067 \lambda$. The internal fine structure of these energy levels is shown in Fig.~\ref{fig:fig4a}. 
Although we expect ground state entanglement to be generated in other multilevel systems, we can isolate just four levels by off-resonantly driving the  $5s4d\,^3{\rm D}_3-5s5d\,^3{\rm D}_2 $ transition at $\rm 600~nm$ with a $\sigma^+$ laser.
This induces an AC Stark shift of only the $\ket{^3{\rm D}_3,m_J\leq 1}$ levels which makes them off resonant to the weak Rabi drive $\Omega$.
If we initialize the atoms in the states $\rm \ket{^3{\rm P}_2,m_J=1}$ and $\rm \ket{^3{\rm P}_2,m_J=2}$, the dynamics will be restricted to $\ket{^3{\rm D}_3,m_J=2}$ and $\ket{^3{\rm D}_3,m_J=3}$, as shown in the shaded region of Fig.~\ref{fig:fig4a}. Only applying Stark shifts to the excited state manifold is enough to isolate the four-level model.

Note that while the state $\rm ^3{\rm P}_2$ is only metastable, it has a long natural lifetime $>100~s$ \cite{yasudaLifetimeMeasurementMetastable2004,porsevHyperfineQuenchingMetastable2004}.
Using this level structure we find that the entanglement generated for this $\rm ^{88}Sr$ case ($N=2$) looks qualitatively similar to our previous four-level results (Figs.~\ref{fig:fig1d},~\ref{fig:fig2f}).

\emph{Conclusions}---
We have shown that the generation of entanglement in the ground state manifold is ubiquitious in weakly-driven multilevel atoms.
This is remarkably different from well-studied weakly-driven two-level atoms, which behave classically. We quantify the entanglement in such driven-dissipative systems via the logarithmic negativity and the more-accessible spin-squeezing of spin-wave modes, which is measurable in current experimental setups for large $N$.
Since the entanglement is produced in the ground subspace, it is long-lived and could potentially be used for the exploration of many-body physics using neutral atoms such as ${}^{88}\text{Sr}$.
Our results open up a number of avenues to explore, such as the effect of entanglement on the light-emission properties, the exploration of new driven-dissipative entangled phases of matter, and the engineering of effective models with non-trivial interactions and dissipation in long-lived subspaces.

\begin{acknowledgments}
\emph{Acknowledgements}---We thank Stefan Lannig and Yongju Hai for their careful reading and helpful comments on the manuscript. This work is supported by the VBFF,   the  NSF JILA-PFC PHY-2317149 and OMA NSF QLCI-2016244 grants, by the DOE Quantum Systems Accelerator (QSA) grant and by NIST.
\end{acknowledgments}

\bibliography{refs}

\clearpage
\widetext
\begin{center}
\textbf{\large Supplementary Material for ``Entanglement generation in weakly-driven arrays of multilevel atoms via dipolar interactions''} \\

\end{center}

\setcounter{equation}{0}
\setcounter{section}{1}
\setcounter{subsection}{1}
\setcounter{subsubsection}{1}
\setcounter{figure}{0}
\setcounter{table}{0}
\setcounter{page}{1}
\makeatletter
\renewcommand{\theequation}{S\arabic{equation}}
\renewcommand{\thefigure}{S\arabic{figure}}
\renewcommand{\thepage}{S\arabic{page}}
\renewcommand{\thesection}{S\arabic{section}}
\renewcommand{\thetable}{S\arabic{table}}
\renewcommand{\thefigure}{S\arabic{figure}}

\date{\today}

\maketitle

\section{Effective ground state model}

\begin{figure}[ht!]
\includegraphics[width=0.35\linewidth]{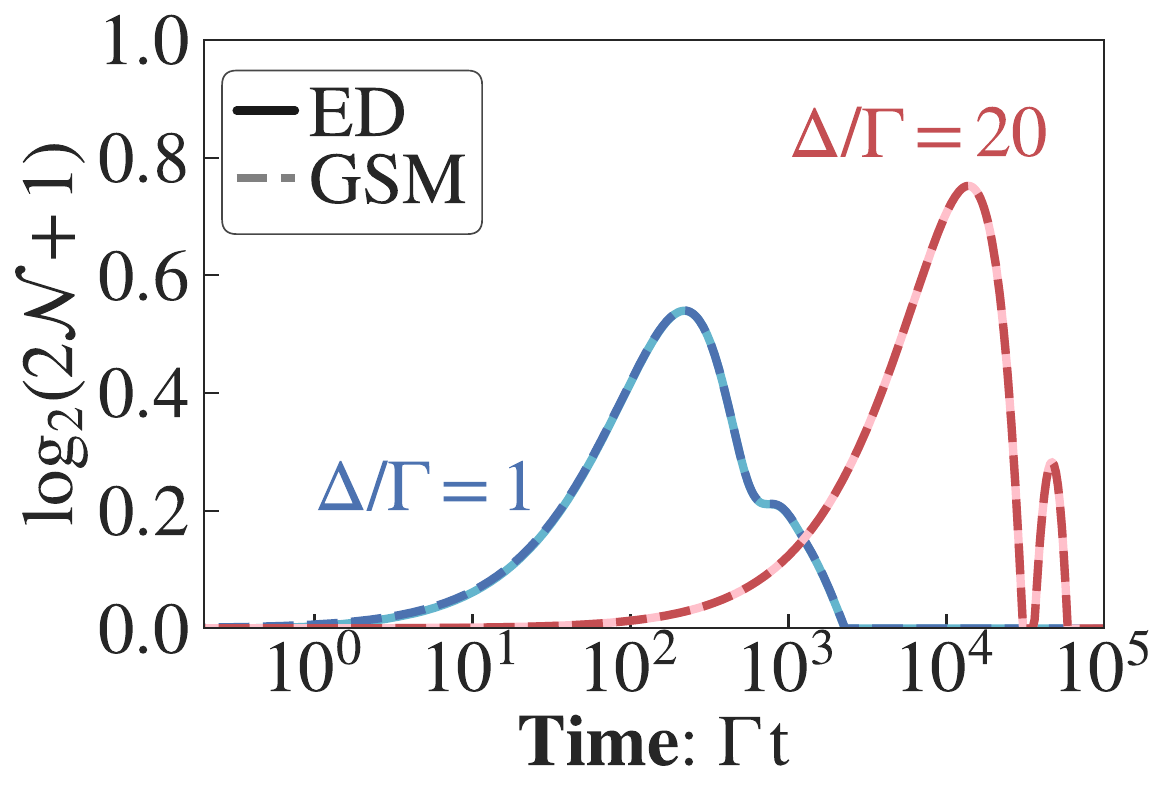}
\caption{\label{fig:GSM_ED_agreement} Agreement between GSM (dashed) and ED (solid) dynamics for $\Omega=0.1\Gamma, r=0.1\lambda, N=2, \hat{e}_{\rm L} = \hat{z}, \ket{\Psi_0}=\left[(\ket{g_-}+\ket{g_+})/\sqrt{2}\right]^{\otimes 2}$.}
\end{figure}

In the weak-driving regime ($\Omega/\Gamma \ll 1$), the excited states are weakly coupled to the ground states via the laser drive. When the system is initialised in the ground state subspace, the population of excited states is suppressed to $\mathcal{O}(\Omega^2/\Gamma^2)$ and the ground state subspace has rich dynamics that lead to strong correlations between the atoms. These correlations are contained in the ground state subspace and can be described by adiabatically eliminating the excited states using second-order perturbation theory. Hence, the system can be described using the effective ground state subspace master equation, $\dot{\hat{\rho}} = -i[\hat{H}_{\rm eff},\hat{\rho}] + \mathcal{L}_{\rm eff}(\hat{\rho})$, where $\hat{H}_{\rm eff}$ is the effective Hamiltonian and $\mathcal{L}_{\rm eff}$ is the effective Lindbladian. $\hat{H}_{\rm eff}$ can be expressed as \cite{reiterEffectiveOperatorFormalism2012a}
\begin{align}
    & \hat{H}_{\rm eff} = -\frac{1}{2} \hat  V_- \left[ \hat H_{\rm NH}^{-1} + \left(\hat H_{\rm NH}^{-1}\right)^\dagger \right]\hat  V_+,\\
    & \hat V_+ = -\sum_{i,q} \Omega_q^i \hat{\mathcal{D}}_q^{i^+} ,\,\, \hat V_-=(\hat V_+)^\dagger,\\
    & \hat H_{\rm NH} = -\Delta \sum_{i,m} \hat{\sigma}_{e_me_m}^i - \sum_{\substack{i,j,\\q,q'}} \left( \Delta_{q,q'}^{ij} + i \Gamma_{q,q'}^{ij} \right)\hat{\mathcal{D}}_q^{i^{+}}\hat{\mathcal{D}}_{q'}^{j^{-}},
\end{align}
where $\Omega_q^j = \Omega e^{i \mathbf{k}\cdot \mathbf{r}_j}$ is the Rabi coupling to atom $j$, $\hat V_+$ is the perturbative single particle laser drive, $\Delta$ is the laser detuning from the single atom transition, and $\hat H_{\rm NH}$ is the non-hermitian Hamiltonian acting on the excited state subspace. Similar to the spin raising  operator in  2-level atoms, we have defined a generalized multilevel raising operator in an atom $i$ as $\hat{\mathcal{D}}_q^{i^{+}} = \sum_{n} C_n^{n+q} \hat{\sigma}_{e_{n+q} g_n}^i$ ($C_n^q$ are Clebsch-Gordan coefficients), which is a superposition of all dipole-allowed transitions through the absorption of a photon of polarisation $\hat{e}_{q=0,\pm 1}$. Furthermore, $\Delta_{q,q'}^{ij} = \hat{e}_q^{*^{\rm T}}\cdot {\rm Re}\mathit{G}(\mathbf{r}_{ij})\cdot\hat{e}_{q'}$ and $\Gamma_{q,q'}^{ij} = \hat{e}_q^{*^{\rm T}}\cdot{\rm Im}\mathit{G}(\mathbf{r}_{ij})\cdot\hat{e}_{q'}$ characterize the elastic and inelastic components of the dipolar interaction, respectively, where $\mathit{G}(\mathbf{r})$ is the vacuum electromagnetic mode's Green's function. $\mathcal{L}_{\rm eff}(\hat{\rho})$ can be expressed as \cite{reiterEffectiveOperatorFormalism2012a}
\begin{align}
    & \mathcal{L}_{\rm eff}(\hat{\rho}) = \sum_{ij,qq'} \Gamma_{q,q'}^{ij} \left( 2\hat{L}_{q'}^{j^-}\hat{\rho} \hat{L}_q^{i^+} -\{\hat{L}_q^{i^+} \hat{L}_{q'}^{j^-},\hat{\rho}\} \right),\\
    & \hat{L}_q^{i^-} = \hat{\mathcal{D}}_q^{i^-}\hat  H_{\rm NH}^{-1}\hat  V_+,\,\,\, \hat{L}_q^{i^+} = (\hat{L}_q^{i^-})^\dagger
\end{align}
where $\hat{L}_q^{i^-}$ are the effective jump operators acting on the ground state subspace. 
In the weak-driving regime, the agreement between the full master equation dynamics (ED) and the ground state manifold master equation dynamics (GSM) is excellent, as shown in Fig.~\ref{fig:GSM_ED_agreement} for a system of $N=2$ atoms with laser polarisation $\hat{e}_{\rm L}=\hat{z}$, inter-atomic spacing $r=0.1\lambda$, and driving strength $\Omega=0.1\Gamma$. We have checked that the agreement between ED and GSM holds in general for other values of $N$, $\hat{e}_{\rm L}$, $r/\lambda$, $\Delta/\Gamma$, and $\Omega/\Gamma \leq 0.1$. Thus, the GSM master equation accurately describes the dynamics of the system in the weak-driving regime.

\section{Entanglement in finite-$N$ 1D arrays}

\subsection{Entanglement dynamics for $N=2$}

To understand the main processes generating the ground state entanglement, we consider $N=2$.
Note that since we are eliminating the excited state, the ground-excited entanglement of regime I is negligible and all the entanglement is between the ground states, i.e., regime II (see main text for definition of regimes).

For the simple case of $N=2$ atoms, $\hat{H}_{\rm eff}$ can be expressed in the Pauli spin basis as
\begin{align}\label{eq:H_eff_N_2_with_coeffs}
    \hat{H}_{\rm eff} &= C_{II} I + \sum_{\alpha,i} C_{\alpha}\hat{\sigma}_\alpha^i + C_{++} (\hat{\sigma}_+^1 \hat{\sigma}_+^2 + h.c.) +  C_{+-} (\hat{\sigma}_+^1 \hat{\sigma}_-^2 + h.c.) + C_{zz} \hat{\sigma}_z^1 \hat{\sigma}_z^2 ,
\end{align}
where $C_{\alpha\beta}$ are obtained from second-order perturbation theory ($\Omega/\Gamma\ll 1$) as a function of the inter-atomic spacing $r/\lambda$ and the laser detuning $\Delta/\Gamma$. 

\subsubsection{Unitary dynamics and the role of $\mathcal{L}_{\rm eff}(\hat{\rho})$}

\begin{figure}[ht!]
\includegraphics[width=0.4\linewidth]{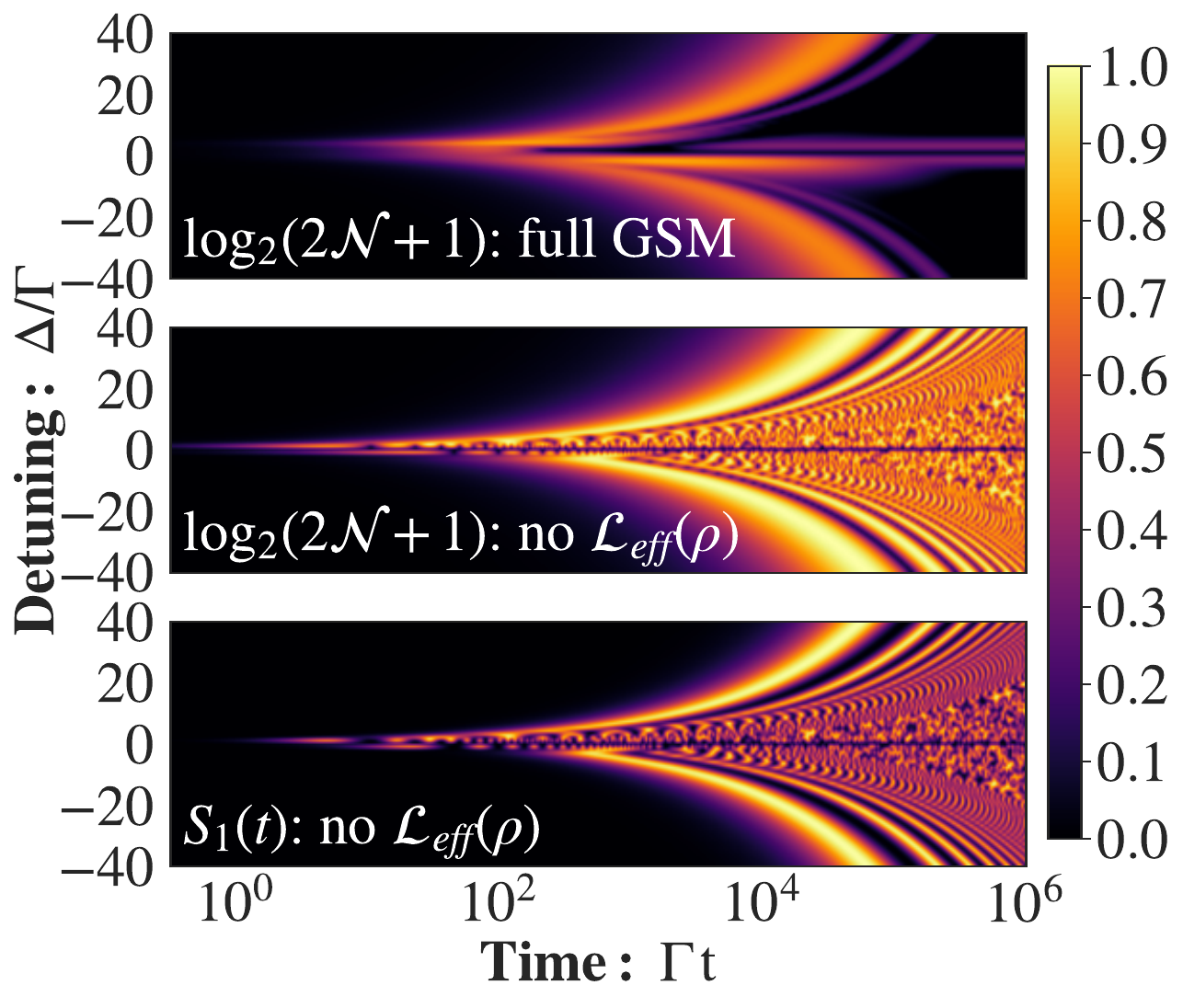}
\caption{\label{fig:compare_numerical_analytical_neg_entropy} $N=2$ entanglement dynamics over a range of detunings. Top: Dynamics of the logarithmic negativity obtained using the full ground-state master equation (full GSM). Middle: Dynamics of the logarithmic negativity obtained using only $\hat{H}_{\rm eff}$ (unitary dynamics) of the ground-state model. Bottom: Dynamics of the Renyi Entropy obtained using only $\hat{H}_{\rm eff}$ (unitary dynamics) of the ground-state model.}
\end{figure}

For $\hat{e}_{\rm L}=\hat{z}$, the single particle Stark shift terms in $\hat{H}_{\rm eff}$ cancel out due to the symmetry of Clebsch-Gordan coefficients for the four-level atom. 
We initialise the system in a product state $\ket{\Psi_0} = (\ket{g_-}+\ket{g_+})^{\otimes 2}/2$. 
The Bell states, $\ket{\lambda_1} = (\ket{g_-g_-}+\ket{g_+g_+})/\sqrt{2}$ and $\ket{\lambda_2} = (\ket{g_-g_+}+\ket{g_+g_-})/\sqrt{2}$, are the eigenstates of $\hat{H}_{\rm eff}$ with eigenvalues $\lambda_1 = C_{zz} + C_{++} + C_{II}$ and $\lambda_2 = -C_{zz} + C_{+-} + C_{II}$. The unitary dynamics of the state is obtained as $\ket{\Psi (t)} =  (\ket{\lambda_1} e^{-i \lambda_1 t} + \ket{\lambda_2} e^{-i \lambda_2 t} ) / \sqrt{2}$ and the second-order Renyi entanglement entropy is  $S_1(t)= -\log_2 \left[0.75 + 0.25 \cos(2(\lambda_1 - \lambda_2)t)\right]$. 
In Fig.~\ref{fig:compare_numerical_analytical_neg_entropy}, we compare the log negativity obtained from the full GSM master equation (including $\mathcal{L}_{\rm eff}(\hat{\rho})$) with the analytical prediction of the log negativity and the Renyi entanglement entropy $S_1(t)$ under purely unitary dynamics. The initial peak entanglement for the two cases is qualitatively similar for large detunings ($\Delta/\Gamma \gg 1$). However, at long-times, the pure state entanglement saturates unlike the dissipative case as $\mathcal{L}_{\rm eff}(\hat{\rho})$ destroys the entanglement for most detunings. 

\subsubsection{Role of $\hat{H}_{\rm eff}$ in generating entanglement}

\begin{figure}[ht!]
\includegraphics[width=0.35\linewidth]{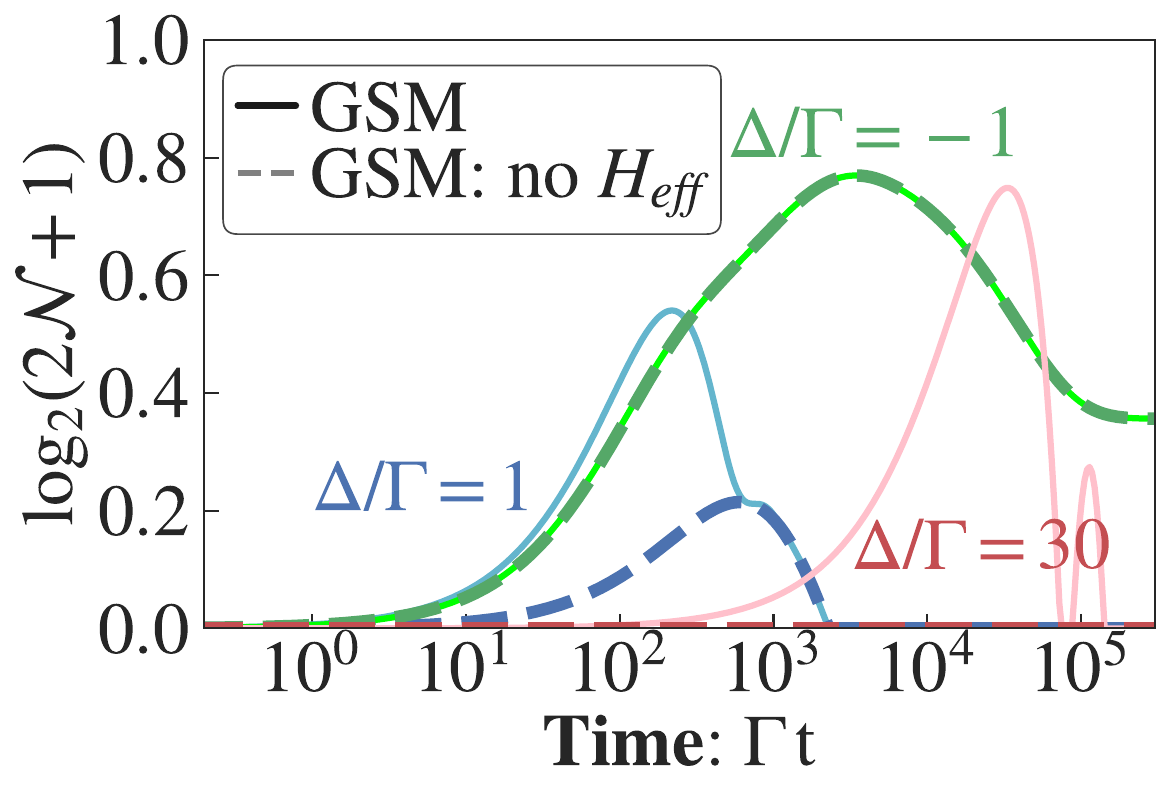}
\caption{ Dynamics of log negativity for $N=2, \hat{e}_{\rm L} = \hat{z}, \ket{\Psi_0}=\left[(\ket{g_-}+\ket{g_+})/\sqrt{2}\right]^{\otimes 2}$; calculated using full ground-state manifold (GSM) master equation (solid) and by turning off $\hat{H}_{\rm eff}$ (dashed).}\label{fig:fig3}
\end{figure}

As shown in Fig.~\ref{fig:fig3}, close to resonance (green solid), the entanglement is largely generated by the many-body terms in the effective jump operators, as entanglement is produced even when setting $\hat{H}_\text{eff}$ to zero (green dashed). Depending on the detuning, the entanglement has contributions from both $\hat{H}_\text{eff}$ and jump operators  (blue solid, dashed). On the other hand, far off from resonance (red solid), the incoherent part has no contribution in generating entanglement as turning off $\hat{H}_{\rm eff}$ leads to zero entanglement (red dashed), and the entanglement is entirely generated by coherent processes.

\subsection{Entanglement dynamics for $N>2$}

\begin{figure}[ht!]
\includegraphics[width=0.4\linewidth]{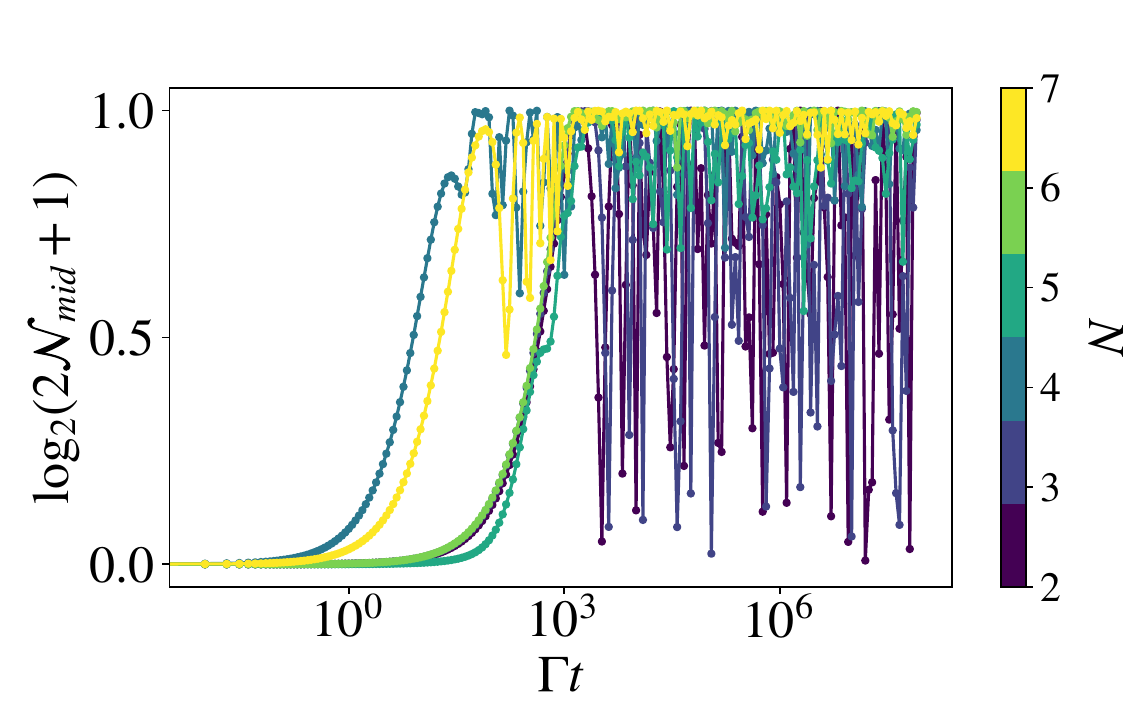}
\caption{\label{fig:neg_no_diss} Dynamics of the logarithmic negativity between the atom in the middle and rest of the array at $\Delta/\Gamma=20$ and $r=0.05\lambda$ for a 1D array over a range of $N$, obtained using ED, under $\hat{H}_{\rm eff}$ (unitary dynamics) of the ground-state model. }
\end{figure}

For $N>2$ as well, the dynamics of the system is governed by $\hat{H}_{\rm eff}$ at early times. The pure state gets thermalized and its logarithmic negativity stays saturated at its maximum value, as shown in Fig.~\ref{fig:neg_no_diss}, for the case of a 1D array in the large detuning regime. We find that the thermalization of the state in the absence of dissipation is actually a generic behavior across all detunings.

\subsection{Other initial conditions and polarizations}
\begin{figure}[ht!]
\includegraphics[width=0.45\linewidth]{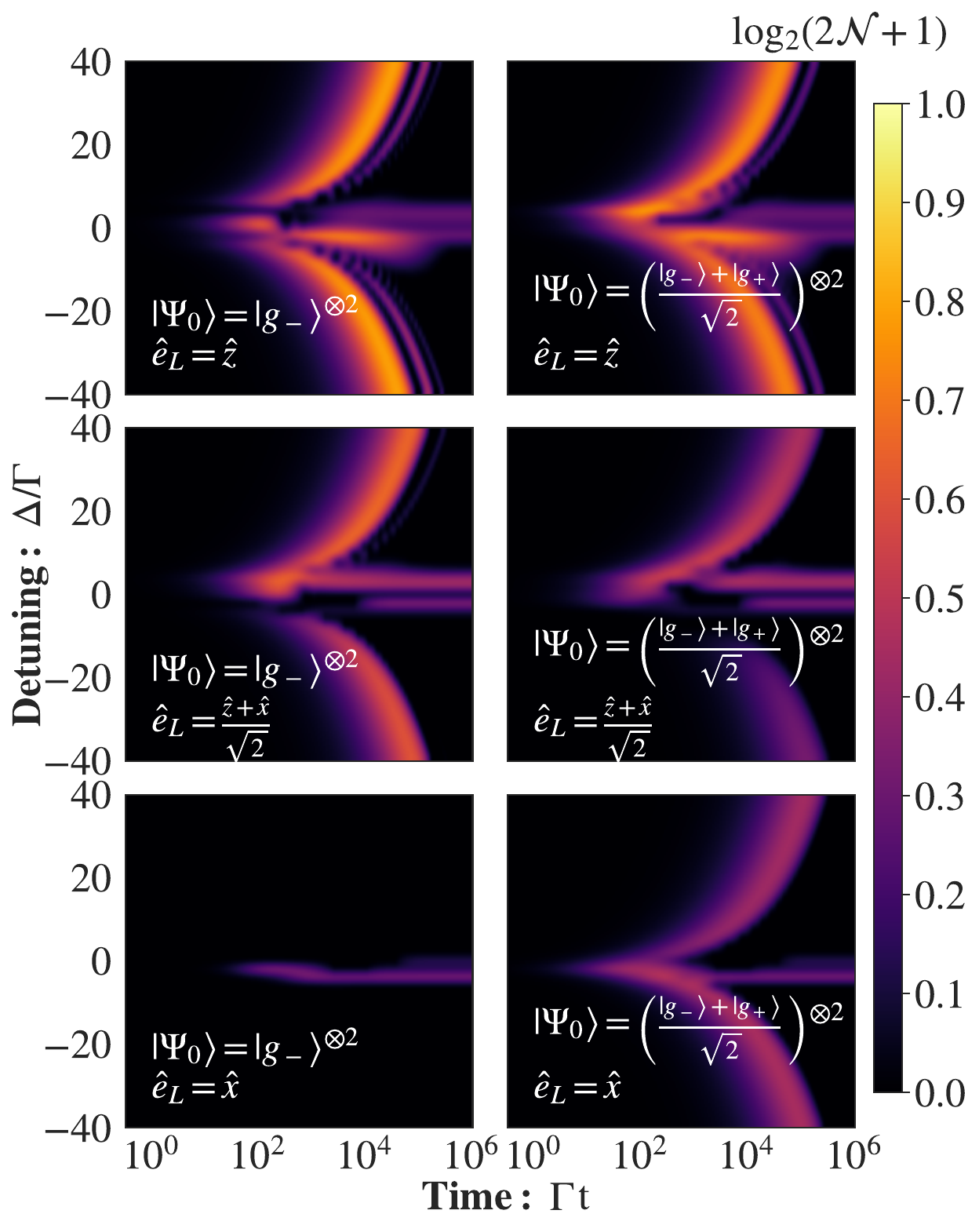}
\caption{\label{fig:initial_state_dependence} Log negativity for $N=2$ for different initial states $\ket{\Psi_0}$ and laser polarisations $\hat{e}_{\rm L}$; calculated using effective ground state master manifold (GSM) master equation.}
\end{figure}

Here, we show how our results generalize to other initial conditions and laser polarizations for the simple case of $N=2$ atoms separated along the $\hat{x}$-axis by a distance $r=0.1\lambda$. As shown in Fig.~\ref{fig:initial_state_dependence}, we find that the generation of negativity is ubiquitous across different configurations but the amount of negativity generated can vary. Overall, the initial condition with an equal superposition of the two ground states produces the largest negativity across polarizations. This is the reason we choose this initial state for most of our analyses throughout our paper. Moreover, we see that the polarization along $\hat{z}$ produces the largest negativity, making it a favorable choice for our work.

\section{Large Detuning Limit}

In the large detuning limit, we treat the interactions perturbatively and expand up to lowest-order as

\begin{align}
    H_{\rm NH}^{-1} &= \frac{1}{-\Delta \sum_{i,m} \hat{\sigma}_{e_me_m}^i - \sum_{\substack{i,j,\\q,q'}} \left( \Delta_{q,q'}^{ij} + i \Gamma_{q,q'}^{ij} \right){\hat {\mathcal D}}^{i^+}_{q} {\hat {\mathcal D}}^{j^-}_{q'} }\approx -\mathbb{P}_e\bigg[\frac{1}{\Delta}- \sum_{\substack{i,j,\\q,q'}} \bigg(\frac{ \mathcal{G}_{q,q'}^{ij}}{\Delta^2}  \bigg){\hat {\mathcal D}}^{i^+}_{q} {\hat {\mathcal D}}^{j^-}_{q'} \bigg] \mathbb{P}_e
\end{align}
where, in the last line we set $\mathcal{G}_{q,q'}^{ij} = \Delta_{q,q'}^{ij} + i \Gamma_{q,q'}^{ij}$ and $\mathbb{P}_e=\sum_{i,m} \hat{\sigma}_{e_me_m}^i$, which is the projection operator onto the excited state subspace. Then, we obtain the effective ground-subspace Hamiltonian as
\begin{align}\label{eq:H_eff_large_det}
    \hat{H}_{\rm eff} &= \mathbb{P}_g \hat{V}_- \mathbb{P}_e\bigg[\frac{1}{\Delta}- \sum_{\substack{i,j,\\q,q'}} \bigg(\frac{ \Delta_{q,q'}^{ij}}{\Delta^2}  \bigg){\hat {\mathcal D}}^{i^+}_{q} {\hat {\mathcal D}}^{j^-}_{q'} \bigg] \mathbb{P}_e\hat{V}_+ \mathbb{P}_g = E_0 - \frac{2\Omega^2}{9\Delta^2} \sum_{i\neq j} \left( 2\Delta_{1,1}^{ij} \hat{\sigma}^+_i \hat{\sigma}^-_j + \Delta_{1,-1}^{ij} \hat{\sigma}^+_i \hat{\sigma}^+_j +  h.c.\right)
\end{align}
where $E_0$ is a constant, we have used the fact that $\Delta_{q,q'}^{ij},\Gamma_
{q,q'}^{ij}$ are real in our geometry, $\mathbb{P}_g=\sum_{i,n} \hat{\sigma}_{g_ng_n}^i$ is the projection operator onto the ground-state subspace, $\hat{\sigma}^+=|g_{+}\rangle\langle g_-|$, and $\hat{\sigma}^-=|g_{-}\rangle\langle g_+|$. The single-body terms in our Hamiltonian cancel out due to the symmetry of the CG coefficients in the four-level system. Similarly, we obtain the effective jump operators as
\begin{align}
    {\hat L}^{j^-}_{0} = \frac{\Omega}{3\Delta} \hat{1}, \,\, {\hat L}^{j^-}_{\pm 1} = \frac{-\sqrt{2}\Omega}{3\Delta} {\hat{\sigma}}_\mp^j .
\end{align}

\section{Spin-squeezing with and without dissipation}

\begin{figure}[ht!]
\includegraphics[width=0.4\linewidth]{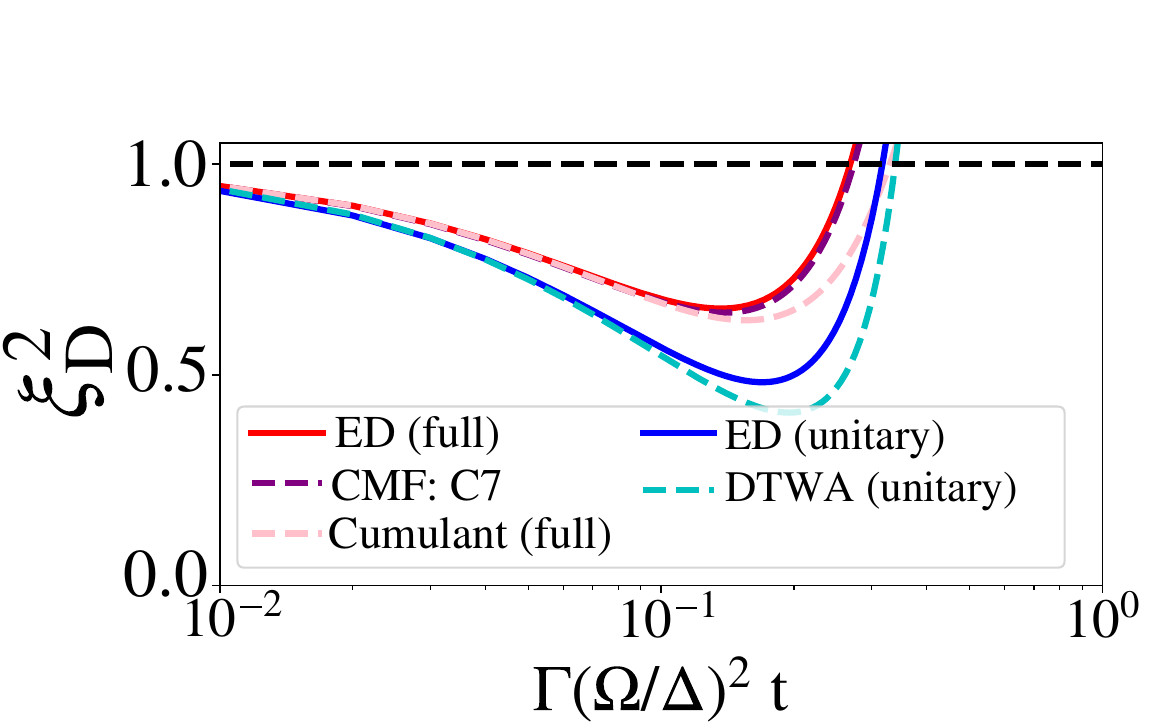} 
\caption{Dynamics of the spin-squeezing parameter ($y$-axis) in (Eq.~\ref{eq:Toth_squeezing}) \cite{TothPRA2009} for a 1D chain of $N=9$ atoms with spacing $r=0.1\lambda$ in the large-detuning limit using ED (solid) with dissipation (red) and only unitary dynamics (blue), cumulant with dissipation (dashed pink), DTWA unitary dynamics (dashed blue), MACE-MF \cite{HazzardPRL2014} (denoted as CMF above, with cluster-size=7) with dissipation (dashed purple).}\label{fig:figsdiss}
\end{figure}

In Fig.~\ref{fig:figsdiss}, we show the dynamics of one type of  spin squeezing parameter, which corresponds to the $\vec{k}=0$ mode, as introduced  in  Ref.~\cite{TothPRA2009} and defined as 
\begin{align}\label{eq:Toth_squeezing}
    \xi^2_{\rm D} = \frac{\lambda_{\rm min}(\chi)}{\langle \mathbf{\hat S}^2 \rangle-N/2}
\end{align}
where $\lambda_{\rm min}(\chi)$ is the smallest eigenvalue of $\chi$, $\chi = (N-1)\mathbf{M} + \mathbf{J}$, $J_{\alpha\beta}=\langle \hat{S}_\alpha\hat{S}_{\beta} + \hat{S}_{\beta}\hat{S}_\alpha\rangle/2$ is the correlation matrix, $M_{\alpha\beta}=J_{\alpha\beta}-\langle \hat{S}_\alpha \rangle \langle\hat{S}_{\beta} \rangle$ is the covariance matrix, $\hat{S}_\alpha = \sum_i \hat{\sigma}_i^\alpha/2$ is the collective spin-operator. We use it to study  a 1D chain in the large detuning limit. As shown in Ref.~\cite{TothPRA2009}, this squeezing parameter  is a valid entanglement witness for mixed states. 

In Fig.~\ref{fig:figsdiss}, at early times, the full dynamics using ED (solid red), which includes dissipation, agrees with the purely unitary dynamics (ED, solid blue), where dissipation is turned off. As time increases, the two curves begin to disagree and the maximal squeezing achieved by the purely unitary dynamics is higher than that obtained from the full dynamics, which implies that dissipation reduces the entanglement in our system. Furthermore, it can be seen that the cumulant method (dashed pink) 
agrees really well with ED, when dissipation is included. For purely unitary dynamics, DTWA (dashed blue) also agrees really well with ED.

\section{Spin-wave Analysis (SWA)}

The initial state of the atomic array is the spin-polarised coherent state pointing along the $\hat{S}_x$-direction on the collective Bloch sphere of $N$-atoms, i.e., $(|g_-\rangle + |g_+\rangle)^{\otimes N}/\sqrt{2^N} \equiv |+\rangle^{\otimes N}$. Then, we define $|+\rangle^{\otimes N}$ as the vacuum of a set of bosonic modes $\{\hat{b}_i\}$ ($i=1\dots N$), such that $\hat{b}_i|+\rangle^{\otimes N} = 0$ \cite{RoscildePRB2023}. Then, we define a new basis of Pauli operators as $\hat{\Sigma}^{z}_i = \hat{\sigma}^x_i,\, \hat{\Sigma}^{y}_i=\hat{\sigma}^y_i,\, \hat{\Sigma}^{x}_i=-\hat{\sigma}^{z}_i$. The new Pauli operators preserve the same commutation relations as the old ones, i.e., $[\hat{\Sigma}^{\alpha}_i,\hat{\Sigma}^{\beta}_j] = 2i\varepsilon_{\alpha\beta\gamma} \hat{\Sigma}^{\gamma}_i \delta_{ij}$, where $\alpha,\beta,\gamma \in \{x,y,z\}$ and $\varepsilon_{\alpha\beta\gamma}$ is the Levi-Civita symbol. In this basis, the effective ground-subspace Hamiltonian in the large-detuning limit (anisotropic XY model) can be expressed as
\begin{align}
    \hat{H}_{\rm eff} &= -\frac{\Omega^2}{9\Delta^2} \sum_{i\neq j} \bigg[ \Delta^{ij}_{1,1} \left(\hat{\Sigma}^{z}_i\hat{\Sigma}^{z}_j + \hat{\Sigma}^{y}_i\hat{\Sigma}^{y}_j\right) + \Delta^{ij}_{1,-1} \left(\hat{\Sigma}^{z}_i\hat{\Sigma}^{z}_j - \hat{\Sigma}^{y}_i\hat{\Sigma}^{y}_j\right) \bigg] .\label{eq:Heff_primed_xyz}
\end{align}

Now, we use the spin-boson mapping under the Holstein-Primakoff approximation \cite{HP1940} as
\begin{align}
    \hat{\Sigma}^{z}_i &= 2(S - \hat{n}_i) = 1 - 2 \hat{n}_i,\,\,\,\hat{\Sigma}^{+}_i = \hat{b}_i \sqrt{2S-\hat{n}_i} \approx \hat{b}_i ,\,\,\,
    \hat{\Sigma}^{-}_i = \hat{b}_i^\dagger \sqrt{2S-\hat{n}_i} \approx \hat{b}_i^\dagger 
\end{align}
where, $S=1/2$ is the spin of each atom, $\hat{n}_i=\hat{b}_i^\dagger \hat{b}_i$ is the number operator and we have assumed that $\hat{n}_i \ll 2S$, i.e., the average occupation of an excited bosonic mode at any site $i$ is very small. We plug $\hat{\Sigma}^{z}_i=1 - 2 \hat{b}_i^\dagger \hat{b}_i$ and $\hat{\Sigma}^{y}_i=-i(\hat{\Sigma}^{+}_i - \hat{\Sigma}^{-}_i) = -i(\hat{b}_i-\hat{b}_i^\dagger)$ into Eq.~(\ref{eq:Heff_primed_xyz}) to express $\hat{H}_{\rm eff}$ in the bosonic basis as
\begin{align}
     \hat{H}_{\rm eff} &= \sum_{i\neq j}  \bigg[ \mathcal{C}_{ij}^x \left(1 - 2\hat{b}_i^\dagger \hat{b}_i - 2\hat{b}_j^\dagger \hat{b}_j\right) + \mathcal{C}_{ij}^y \left(\hat{b}_i^\dagger \hat{b}_j + \hat{b}_i \hat{b}_j^\dagger - \hat{b}_i \hat{b}_j - \hat{b}_i^\dagger \hat{b}_j^\dagger\right)  \bigg],
\end{align}
where we have assumed that non-linear terms such as $\hat{n}_i\hat{n}_j$ are suppressed, $\mathcal{C}_{ij}^x=-\frac{\Omega^2}{9\Delta^2}\left(\Delta^{ij}_{1,1} + \Delta^{ij}_{1,-1} \right)$, and $\mathcal{C}_{ij}^y=-\frac{\Omega^2}{9\Delta^2}\left(\Delta^{ij}_{1,1} - \Delta^{ij}_{1,-1} \right)$. The first term $\left(1 - 2\hat{b}_i^\dagger \hat{b}_i - 2\hat{b}_j^\dagger \hat{b}_j\right)$ is the vacuum+on-site energy term and the second term consists of two parts -- the hopping term $\left(\hat{b}_i^\dagger \hat{b}_j + \hat{b}_i \hat{b}_j^\dagger\right)$ and the pair-creation term $\left( \hat{b}_i \hat{b}_j + \hat{b}_i^\dagger \hat{b}_j^\dagger\right)$. 

We define the momentum-space bosonic operators as
\begin{align}
    \hat{b}_{\vec{k}} &= \frac{1}{\sqrt{N}} \sum_{i=1}^N e^{ -i \vec{k} \cdot \vec{r}_{i}} \hat{b}_i 
\end{align}
where $\vec{r}_j = a ( j\mod \sqrt{N}\hat{X} + \lfloor j/ \sqrt{N} \rfloor\hat{Z})$, $a$ is the lattice spacing, $\sqrt{N}$ is a whole number, $j=1\dots \sqrt{N}$, $\vec{k} = k_X \hat{X} + k_Z \hat{Z} $, where $k_X,k_Z \in \{2n \pi/(\sqrt{N}a),\,\,n=1,\dots,\sqrt{N}\}$. We define the Fourier expansion of the interaction coefficients as $C^y_{ij} \equiv C^y(\vec{r}_{ij}) = (1/N)\sum_{\vec{k}} \exp (i \vec{k} \cdot \vec{r}_{ij}) \tilde{\mathcal{C}}^y_{\vec{k}}$ and similarly, $C^x_{ij}$. Using this, we obtain the momentum-space bosonic Hamiltonian of our system, up to linear order in excitations, as
\begin{align}\label{eq:Heff_bkbkdagger}
\hat{H}_{\rm eff}&=  \sum_{\vec{k}}  \bigg[ \varepsilon_{\vec{k}}  \hat{b}_{\vec{k}}^\dagger  \hat{b}_{\vec{k}} -  \frac{\Omega_{\vec{k}}}{2}\left(  \hat{b}_{\vec{k}}  \hat{b}_{-\vec{k}}  + \hat{b}_{\vec{k}}^\dagger  \hat{b}_{-\vec{k}}^\dagger \right)  \bigg] 
\end{align}
where, $\varepsilon_{\vec{k}}= -  4 \tilde{\mathcal{C}}^x_{\vec{0}} + \tilde{\mathcal{C}}^y_{\vec{k}}   + \tilde{\mathcal{C}}^y_{-\vec{k}}$ is the hopping coefficient and $\Omega_{\vec{k}} = 2 \tilde{\mathcal{C}}^y_{\vec{k}}$ is the pair-creation coefficient. $\varepsilon_{\vec{k}} $ and $\Omega_{\vec{k}} $ are real because $\Delta_{1,1}^{ij},\Delta_{-1,1}^{ij} \in \mathbb{R}$ in our chosen geometry and $\Delta_{q,q'}^{ij} = \Delta_{q,q'}^{ji}$. 

To obtain the eigenmodes of $\hat{H}_{\rm eff}$ (Eq.~(\ref{eq:Heff_bkbkdagger})), we define new bosonic operators using the Bogoliubov transformation \cite{Valatin1958CommentsOT,Bogoljubov1958OnAN} as
\begin{align}
    \hat{a}_{\vec{k}} & = u(\vec{k}) \hat{b}_{\vec{k}} - v(\vec{k})^* \hat{b}_{-\vec{k}}^\dagger, 
\end{align}
and we obtain $u(\vec{k})=\cosh\theta(\vec{k})e^{i \alpha(\vec{k})}$, $v(\vec{k})=\sinh\theta(\vec{k})e^{i \beta(\vec{k})}$, $\theta(\vec{k}) = \theta(-\vec{k})$, and $\alpha(\vec{k})-\beta(-\vec{k}) = \alpha(-\vec{k})-\beta(\vec{k}) + 2n\pi$ by preserving the bosonic commutation relations. Moreover, by satisfying the eigenvalue equation, we obtain 
\begin{align}
    \cosh(2\theta(\vec{k})) = \frac{\varepsilon_{\vec{k}}}{\xi_{\vec{k}}},\,\,\, \sinh(2\theta(\vec{k})) = \frac{\Omega_{\vec{k}}}{\xi_{\vec{k}}}\cos(\alpha(\vec{k}) + \beta(\vec{k}))
\end{align}
with $\sin(\alpha(\vec{k}) + \beta(\vec{k})) = 0$ and $\xi_{\vec{k}} = \sqrt{\varepsilon_{\vec{k}}^2-\Omega_{\vec{k}}^2}$. Then, we can rewrite the effective Hamiltonian as $\hat{H}_{\rm eff}= \sum_{\vec{k}}\left\{ \hat{a}_{\vec{k}}^\dagger \hat{a}_{\vec{k}} \left(\frac{\xi_{\vec{k}} }{2} + \frac{\varepsilon_{\vec{k}} }{2}\right) + \hat{a}_{-\vec{k}}^\dagger \hat{a}_{-\vec{k}} \left(\frac{\xi_{\vec{k}} }{2} - \frac{\varepsilon_{\vec{k}} }{2} \right) \right\}$. Without loss of generality, we replace $-\vec{k}\to \vec{k}$ as the sum includes both the positive and the corresponding negative momenta, and get
\begin{align}
    \hat{H}_{\rm eff} =  \sum_{\vec{k}} \xi_{\vec{k}} \hat{a}_{\vec{k}}^\dagger \hat{a}_{\vec{k}}  .
\end{align}
The eigenmodes evolve in time as $\hat{a}_{\vec{k}}(t) = e^{-i\xi_{\vec{k}}t} \hat{a}_{\vec{k}}(0) $. Using, this we obtain the dynamics of the original bosonic modes as
\begin{align}
    \hat{b}_{\vec{k}}(t) & =  \hat{b}_{\vec{k}}(0) \left( \cos(\xi_{\vec{k}}t) - i \sin(\xi_{\vec{k}}t) \cosh(2\theta_{\vec{k}}) \right) + i \hat{b}_{-\vec{k}}^\dagger(0) e^{-i(\alpha(\vec{k})+\beta(\vec{k}))} \sin(\xi_{\vec{k}}t) \sinh(2\theta_{\vec{k}}). 
\end{align}
and the dynamics of the mode occupation as
\begin{align}
    \langle \hat{n}_{\vec{k}} (t) \rangle &= \langle 0| \hat{b}_{\vec{k}}^\dagger(t)\hat{b}_{\vec{k}}(t) |0\rangle= \frac{|\Omega_{\vec{k}}|^2}{\xi_{\vec{k}}^2} \sin^2(\xi_{\vec{k}} t)
\end{align}
for $\xi_{\vec{k}}^2>0$, i.e., $|\varepsilon_{\vec{k}}|>|\Omega_{\vec{k}}|$, and as
\begin{align}
    \langle \hat{n}_{\vec{k}} (t) \rangle &= \frac{|\Omega_{\vec{k}}|^2}{|\xi_{\vec{k}}|^2} \sinh^2(|\xi_{\vec{k}}| t)
\end{align}
for $|\varepsilon_{\vec{k}}|<|\Omega_{\vec{k}}|$. We have $\xi_{\vec{k}}^2 =  16 \tilde{\mathcal{C}}^x_{\vec{0}} ( \tilde{\mathcal{C}}^x_{\vec{0}}-\tilde{\mathcal{C}}^y_{\vec{k}})$ so, in the absence of the pair-creation term, i.e., $\tilde{\mathcal{C}}^y_{\vec{k}}=0$, we have $\xi_{\vec{k}}^2>0$ for all $\vec{k}$, and no exponential growth of modes. 

\begin{figure}[ht!]
\includegraphics[width=0.35\linewidth]{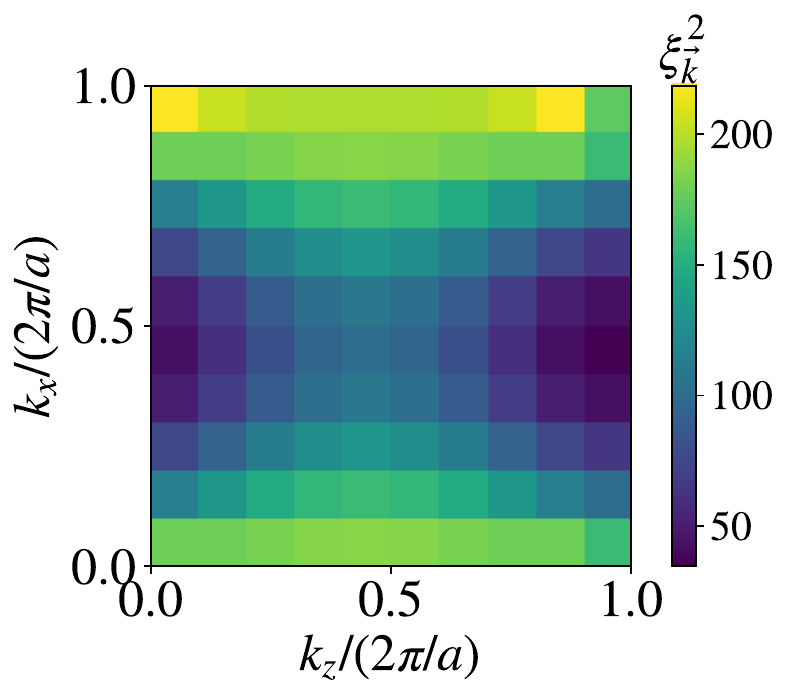} 
\caption{Spectrum $\xi_{\vec{k}}^2$ of bosonic modes for a $10\times 10$ array in the $X-Z$ plane with lattice spacing $a=0.1\lambda$ and polarization $\hat{e}_{\rm L}=\hat{e}_0 = \hat{Z}$, obtained from spin-wave analysis.}\label{fig:xisqr_N_100}
\end{figure}

In Fig.~\ref{fig:xisqr_N_100}, we show the spectrum, $\xi_{\vec{k}}^2$, of modes for a $10\times 10$ array in the $X-Z$ plane with $\hat{e}_{\rm L}=\hat{e}_0 = \hat{Z}$. We see that the spectrum is positive for all $\vec{k}$, such that there is no unstable mode. However, we show in the main text that the lowest spin-wave energy (lowest $\xi_{\vec{k}}^2$) modes have the highest growth in the mode occupation, $\langle \hat{n}_{\vec{k}} (t) \rangle$, which is also known as the spin-structure factor. These modes have relatively large $|\Omega_{\vec{k}}| \equiv 2|\tilde{\mathcal{C}}^y_{\vec{k}}|$, which is the pair-creation coefficient.

The bosonic quadratures in position space are defined as $\hat{X}_i = (\hat{b}_i + \hat{b}_i^\dagger)/\sqrt{2} \equiv \hat{\Sigma}^{x}_i /\sqrt{2} $ and $\hat{P}_i = -i(\hat{b}_i - \hat{b}_i^\dagger)/\sqrt{2} \equiv \hat{\Sigma}^{y}_i /\sqrt{2} $. Then, the commutation relation is given as $[\hat{X}_i,\hat{P}_j] = i \delta_{ij}$, which leads to the uncertainty relation, $\langle|\Delta \hat{X}_i|^2\rangle \langle|\Delta \hat{P}_j|^2\rangle \geq 1/4$. We define the momentum-space quadratures as
\begin{align}
    \hat{X}_{\vec{k}} = \frac{(\hat{b}_{\vec{k}} + \hat{b}_{\vec{k}}^\dagger)}{\sqrt{2}}, \,\,\, \hat{P}_{\vec{k}} = \frac{-i(\hat{b}_{\vec{k}} - \hat{b}_{\vec{k}}^\dagger)}{\sqrt{2}}
\end{align}
such that the commutation relation is preserved, $[\hat{X}_{\vec{k}},\hat{P}_{\vec{k}'}] = i \delta_{{\vec{k}},{\vec{k}'}}$. It can be seen that $\hat{X}_{\vec{k}}^\dagger=\hat{X}_{\vec{k}}$ and $\hat{P}_{\vec{k}}^\dagger=\hat{P}_{\vec{k}}$. Now, we define generalized quadratures $\hat{Q}_{\vec{k}}=\hat{X}_{\vec{k}}\cos\phi+\hat{P}_{\vec{k}}\sin\phi$ and $\hat{R}_{\vec{k}}=-\hat{X}_{\vec{k}}\sin\phi+\hat{P}_{\vec{k}}\cos\phi$, where $\phi$ is the angle in phase-space from $\hat X_{\vec{k}}$. We obtain the variance as $\langle|\Delta \hat{Q}_{\vec{k}}|^2\rangle=\langle \hat{X}_{\vec{k}}^2\rangle \cos^2\phi + \langle \hat{P}_{\vec{k}}^2\rangle \sin^2\phi + \langle \hat{X}_{\vec{k}} \hat{P}_{\vec{k}}\rangle \sin(2\phi)/2 + \langle \hat{P}_{\vec{k}} \hat{X}_{\vec{k}}\rangle \sin(2\phi)/2$, where we have used $\langle 0| \hat{X}_{\vec{k}}|0 \rangle = 0$ and $\langle 0| \hat{P}_{\vec{k}} | 0\rangle = 0$. Then, we obtain the dynamics of these variances as
\begin{align}\label{eq:Qksqr}
\langle|\Delta \hat{Q}_{\vec{k}} (t,\phi)|^2\rangle&= \frac{1}{2} + \frac{|\Omega_{\vec{k}}|^2}{\xi_{\vec{k}}^2} \sin^2(\xi_{\vec{k}}t) + \bigg[\cos(2\phi)  \frac{\Omega_{\vec{k}}\varepsilon_{\vec{k}}}{\xi_{\vec{k}}^2} \sin^2(\xi_{\vec{k}}t) +\sin(2\phi) \frac{\Omega_{\vec{k}}}{2\xi_{\vec{k}}}  \sin(2\xi_{\vec{k}}t) \bigg] f(\vec{k}) ,
\end{align}
where $f(\vec{k})=\delta_{\vec{k},\vec{0}} + \delta_{k_X,\pi/a} \delta_{k_Z,\pi/a} + \delta_{k_X,\pi/a} \delta_{k_Z,0} + \delta_{k_X,0} \delta_{k_Z,\pi/a}$ and $\langle|\Delta \hat{R}_{\vec{k}} (t,\phi)|^2\rangle = \langle|\Delta \hat{Q}_{\vec{k}} (t,\phi+\pi/2)|^2\rangle$. Then, to diagonalize the covariance matrix, we find $\phi_*$ as the phase-space angle where the variance $\langle|\Delta \hat{Q}_{\vec{k}} (t,\phi)|^2\rangle$ has an extremum. This condition gives us
\begin{align}
    \frac{d \langle|\Delta Q_{\vec{k}}|^2\rangle}{d\phi} = 0 \Rightarrow \tan (2\phi_*) = \frac{\xi_{\vec{k}}}{\varepsilon_{\vec{k}}}\cot(\xi_{\vec{k}}t)
\end{align}
for $f(\vec{k})\neq 0$. For modes with $f(\vec{k})=0$, there is no squeezing and the variance is independent of $\phi$.

\begin{figure}[ht!]
\includegraphics[width=0.35\linewidth]{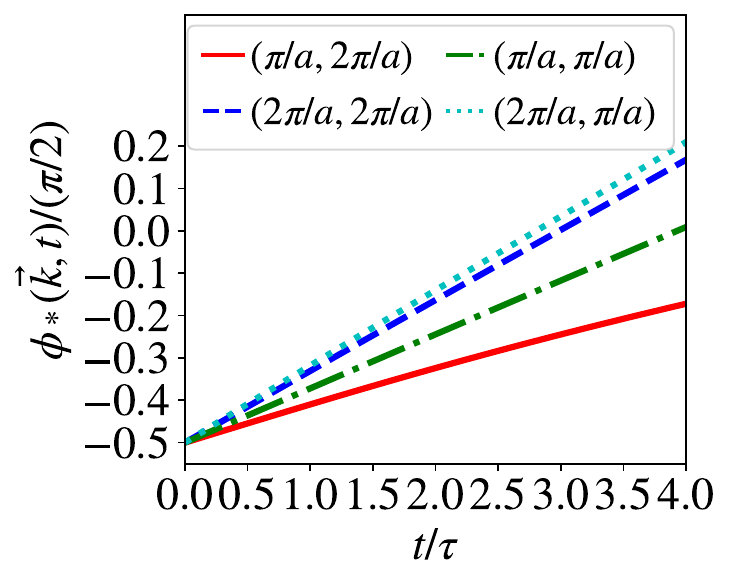}
\caption{\label{fig:figsphi_star} Dynamics of optimum angle $\phi_*(\vec{k},t)$ for $\vec{k}\equiv (k_X,k_Z)$ for a 2D array of $10\times 10$ atoms in the $X-Z$ plane, with $\hat{e}_{\rm L}=\hat{e}_0 = \hat{Z}$.  Time is in units of $\tau=0.04\Delta^2/(\Gamma\Omega^2)$, same as the main text. Data obtained using SWA.}
\end{figure}

As can be seen from above, the variance $\langle|\Delta \hat{Q}_{\vec{k}} (t,\phi)|^2\rangle$ at $\phi=0$ reduces to $\langle \hat{X}_{\vec{k}}^2\rangle$ and at $\phi=\pi/2$ to $\langle \hat{P}_{\vec{k}}^2\rangle$. Furthermore, we can go back to the spin-variables and express these variances as
\begin{align}
    \langle \hat{X}_{\vec{k}}^2\rangle & = \frac{2}{N} \langle \hat{\tilde S}_z^2 ({\vec{k}})\rangle ,\,\,\,
    \langle \hat{P}_{\vec{k}}^2\rangle  = \frac{2}{N} \langle \hat{\tilde S}_y^2 ({\vec{k}}) \rangle
\end{align}
where, same as in the main text, we have defined the spin-wave operators as $\hat{\tilde{S}}_{z} (\vec{k})=-\sum_i (\hat \Sigma_i^{+} e^{i\vec{k}\cdot \vec{r}_i} + \hat \Sigma_i^{-} e^{-i\vec{k}\cdot \vec{r}_i})/2$ and $\hat{\tilde{S}}_{y} (\vec{k})=-i\sum_i (\hat \Sigma_i^{+} e^{i\vec{k}\cdot \vec{r}_i} -\hat \Sigma_i^{-} e^{-i\vec{k}\cdot \vec{r}_i})/2$, with the initial magnetisation along $\hat{{S}}_{x} = \sum_i \hat \sigma_i^{x}/2 \equiv \sum_i \hat \Sigma_i^{z}/2  $. Then, we can define the spin-wave raising (lowering) operators for mode $\vec{k}$ as $\hat{\tilde S}_+ ({\vec{k}}) = -\hat{\tilde S}_z ({\vec{k}}) + i \hat{\tilde S}_y ({\vec{k}}) \,\, \left(\hat{\tilde S}_-({\vec{k}})=\hat{\tilde S}_+^\dagger({\vec{k}})\right) $. The commutation relations for these operators are  $[\hat{\tilde S}_+ ({\vec{k}}),\hat{\tilde S}_- ({\vec{k}})] = 2\hat{S}_x$ and $[\hat{S}_x,\hat{\tilde S}_+ ({\vec{k}})]  = \hat{\tilde S}_+ ({\vec{k}})$. 
Thus, the spin-wave operators for a general mode $\vec{k}$ satisfy the same commutation relations as the ones satisfied by the $\vec{k}=0$ mode, i.e., $[\hat{ S}_+ ,\hat{ S}_- ] = 2 \hat{S}_x $ and $[\hat{ S}_x ,\hat{ S}_+ ] = \hat{ S}_+ $, where $\hat{ S}_\pm = -\hat S_z \pm i \hat S_y$ and $\hat{ S}_\alpha = \sum_i \hat\sigma_i^\alpha /2$, $\alpha=x,y,z$. 
Hence, we treat the spin-wave operators $\hat{\tilde S}_\alpha ({\vec{k}})$ as collective ``observables'' of the system and consider the variances in these operators as the squeezing/anti-squeezing of the spin-wave modes. The variances for mode $\vec{k}=0$ satisfy the uncertainty relation, $\langle\Delta \hat S_z^2 \rangle \langle\Delta \hat S_y^2 \rangle \geq \langle \hat S_x \rangle^2/4$, which leads to the spin-squeezing condition, ${N\langle \Delta\hat{ S}_\alpha^2  \rangle}/{\langle \hat S_x \rangle^2} \leq 1$ for ${\alpha \in \{y,z\}}$ \cite{KitagawaPRA1993,winelandPRA1994}. Similarly, for a general mode $\vec{k}$, we use the commutation relations described above to get the uncertainty relation as $\langle\Delta \hat {\tilde{S}}_z^2({\vec{k}}) \rangle \langle\Delta \hat {\tilde{S}}_y^2({\vec{k}}) \rangle \geq \langle \hat S_x \rangle^2/4$. 
Further, we define a generalized spin-wave operator at an angle $\phi$ in phase-space as $\hat{\tilde{S}}_{\phi}(\vec{k})=-\hat{\tilde{S}}_{z}(\vec{k}) \cos\phi + \hat{\tilde{S}}_{y}(\vec{k}) \sin\phi$. 
Then, analogous to mode $\vec{k}=0$, we define a ``generalized'' spin-squeezing condition for a mode $\vec{k}$ as 
\begin{align}
    \frac{N\langle \Delta \hat{\tilde S}_\phi^2 ({\vec{k}}) \rangle}{\langle \hat S_x \rangle^2} \leq 1,
\end{align}
where the variance further simplifies to $\langle \Delta \hat{\tilde S}_\phi^2 ({\vec{k}}) \rangle = \langle \hat{\tilde S}_\phi^2 ({\vec{k}}) \rangle$ as $\langle \hat{\tilde S}_\phi ({\vec{k}}) \rangle = 0$ at all times for our system.
A recent work \cite{rosario2024detecting} further discusses the validity of the spin-squeezing of spin-wave modes as an entanglement witness. In Fig.~\ref{fig:figsphi_star}, we show the optimum values of $\phi \equiv \phi_*$ at which spin-squeezing is minimized for four modes, $\vec{k} \equiv (k_X,k_Z) = (\pi/a,2\pi,a), (2\pi/a,2\pi,a), (\pi/a,\pi,a), (2\pi/a,\pi,a)$ for a $10\times 10$ array in the $X-Z$ plane with $a=0.1\lambda$, same as the case shown in the main text. We focus on these modes as SWA predicts a growth in squeezing only for these modes in our system, as seen in Eq.~(\ref{eq:Qksqr}).

\section{Tuning the structure of modes with laser polarization}

\begin{figure*}[]
\includegraphics[width=\linewidth]{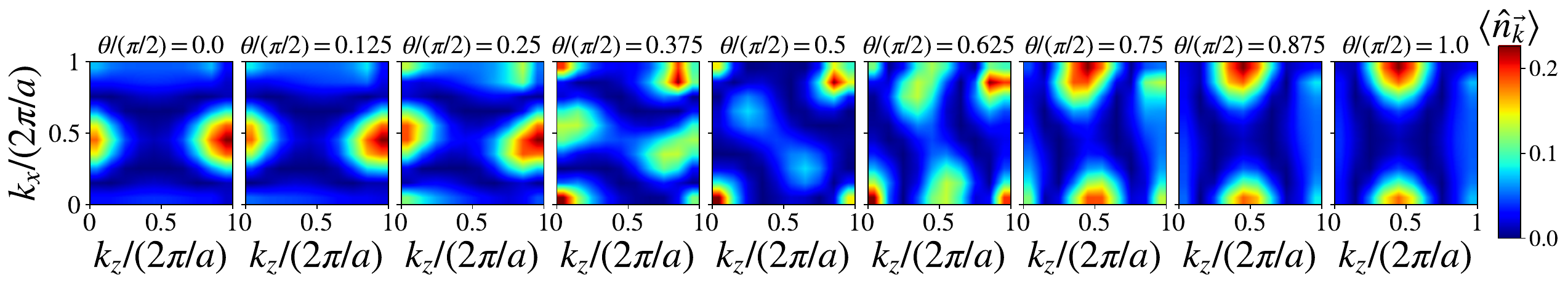} 
\caption{\label{fig:figthetank} Mode occupation (spin-structure factor) of a $10\times 10$ array in the $X-Z$ plane at a fixed time $\Gamma (\Omega/\Delta)^2 t = 0.1$ with $\hat{e}_{\rm L}=\hat{e}_0 = \sin\theta \hat{X} + \cos\theta \hat{Z}$ for a range of $\theta$, obtained from SWA.}
\end{figure*}

\begin{figure*}[]
\includegraphics[width=\linewidth]{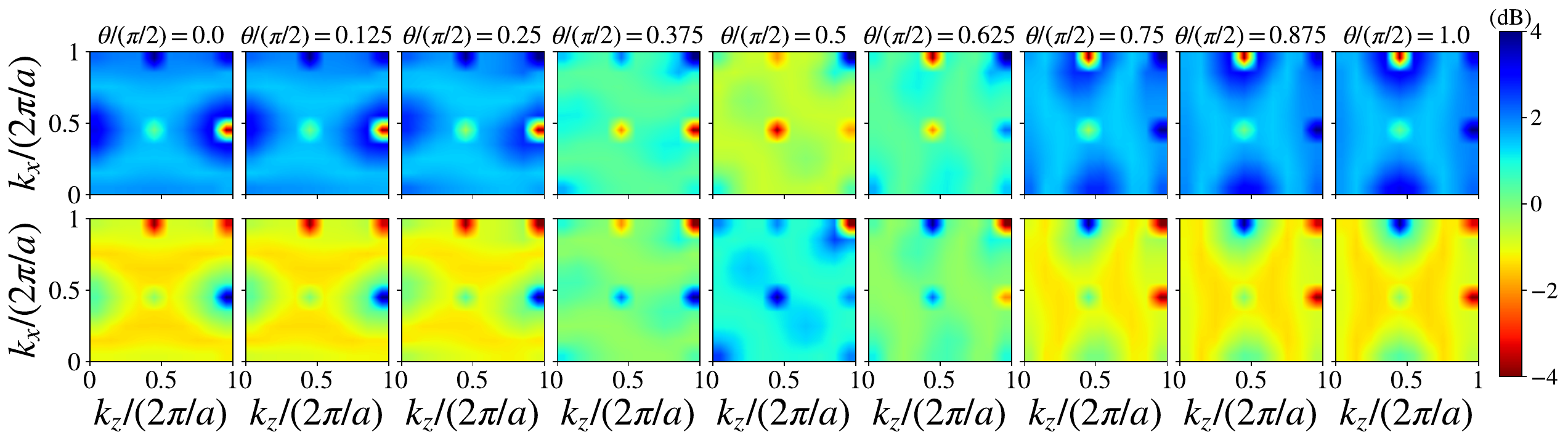} 
\caption{\label{fig:figthetaQksqr} Quadrature variances of a $10\times 10$ array in the $X-Z$ plane at fixed time $\Gamma (\Omega/\Delta)^2 t = 0.1$ with $\hat{e}_{\rm L}=\hat{e}_0 = \sin\theta \hat{X} + \cos\theta \hat{Z}$ for a range of $\theta$, obtained from SWA. Top row: $4\langle \hat{\tilde S}_{\phi_*}^2 ({\vec{k}}) \rangle/N $. Bottom row: $4\langle \hat{\tilde S}_{\phi_*+\pi/2}^2 ({\vec{k}}) \rangle/N $. These variances correspond to the development of squeezing/anti-squeezing, as under the HP approximation, $\langle \hat S_x\rangle = N/2$. Squeezing is optimal at $\phi_*$.}
\end{figure*}

We define an angle $\theta$ between the $Z$-axis and the laser polarization, such that $\hat{e}_{\rm L} = \hat{e}_0 = \sin\theta \hat{X} + \cos\theta \hat{Z}$ and $\hat{e}_\pm = \mp\left((\cos\theta \hat{X} - \sin\theta\hat{Z}) \pm i\hat{Y}\right)\sqrt{2}$. Then, by tuning $\theta$, we can access the growth of different modes as seen in the spin-structure factor in Fig.~\ref{fig:figthetank}, for a $10\times 10$ array at a fixed time $\Gamma (\Omega/\Delta)^2 t = 0.1$ using SWA. The growth of modes changes across $\theta$ as the interaction coefficients also change with $\theta$. 

In Fig.~\ref{fig:figthetaQksqr}, we show how the variances in the quadratures $\langle \hat{Q}_{\vec{k}}^2 (t,\phi_*)\rangle = 4\langle \hat{\tilde S}_{\phi_*}^2 ({\vec{k}}) \rangle/N$ (top row) and $\langle \hat{R}_{\vec{k}}^2 (t,\phi_*)\rangle = 4\langle \hat{\tilde S}_{\phi_*+\pi/2}^2 ({\vec{k}}) \rangle/N$ (bottom row) change with $\theta$ for the same system and time stamp as Fig.~\ref{fig:figthetank}. The mode occupation and the variances show similar behavior for the different modes across $\theta$. Moreover, we do not see exceptional enhancement of the smallest variance across all modes for any $\theta$. The effect of changing $\theta$ can be understood more or less as a rotation of the structure of the modes in the $k_X-k_Z$ plane, without changing the physics qualitatively.

\section{Comparing the structure of modes for 1D and 2D}

\begin{figure}[ht!]
\includegraphics[width=0.3\linewidth]{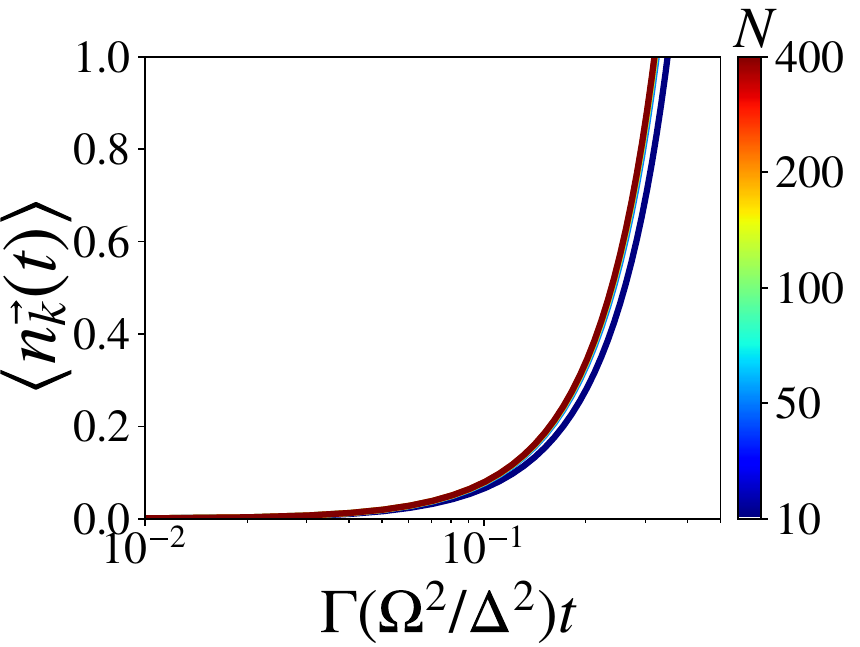}
\includegraphics[width=0.3\linewidth]{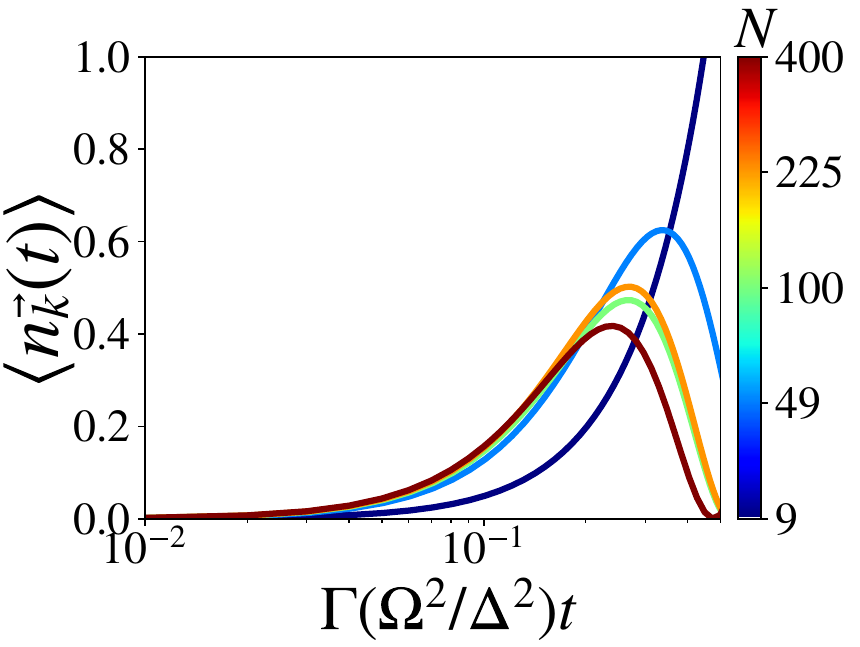}
\caption{\label{fig:figsnk1D2Dkxpi} Dynamics of mode occupation for: (left) mode $\vec{k}=(\pi/a)\hat{X}$ in a 1D array of $N$ atoms along the $X$ axis and (right) mode $\vec{k}=(\pi/a)\hat{X}+(2\pi/a)\hat{Z}$ in a 2D array of $\sqrt{N}\times\sqrt{N}$ atoms in the $X-Z$ plane, with $\hat{e}_{\rm L}=\hat{e}_0 = \hat{Z}$. $a=$ lattice constant. Data obtained using SWA.}
\end{figure}

In Fig.~\ref{fig:figsnk1D2Dkxpi}, we show the dynamics of the mode occupation (spin-structure factor) for the $k_X=\pi/a$ ($a=$ lattice constant) mode for a 1D array with atoms along the $X$-axis and a 2D array. For the 1D array, this mode is unstable and the occupation grows exponentially even at large $N$. For 2D, this mode is unstable at small $N\sim 9$ but becomes stable at large $N$. Despite becoming stable for 2D, this mode still has the lowest energy, as shown earlier in Fig.~\ref{fig:xisqr_N_100}, and has the largest growth in the mode occupation, as shown in the main text. 

To understand the instability of this mode in 1D, we first consider the coefficients of the XY model for our 1D array -- $ \mathcal{C}_{ij}^x = \frac{\Gamma \Omega^2}{12\Delta^2} \frac{\cos (k_0 |{\bf r}_{ij}|)}{\big(k_0 |{\bf r}_{ij}|\big )^3}$ and $\mathcal{C}_{ij}^y = \mathcal{C}_{ij}^x\left(1 - 6({\hat r}_{ij}\cdot\hat{e}_+)^2\right) =-3 \mathcal{C}_{ij}^x$ as $({\hat r}_{ij}\cdot\hat{e}_+)^2=1/2$ for all $i,j$. Then, we obtain the Fourier Transform (FT) of these coefficients as $ \tilde{\mathcal{C}}^{x,y}_{\vec{k}} = \sum_{\vec{r}_{ij}} \exp (-i \vec{k} \cdot \vec{r}_{ij}) \mathcal{C}_{ij}^{x,y}$, and we get $ \tilde{\mathcal{C}}^{y}_{\vec{k}} = -3\tilde{\mathcal{C}}^{x}_{\vec{k}}$. Then, we obtain the hopping coefficient $\varepsilon_{\vec{k}}= -  4 \tilde{\mathcal{C}}^x_{\vec{0}} + \tilde{\mathcal{C}}^y_{\vec{k}}   + \tilde{\mathcal{C}}^y_{-\vec{k}} = -4 \tilde{\mathcal{C}}^x_{\vec{0}} - 6\tilde{\mathcal{C}}^x_{\vec{k}} $  and the pair-creation coefficient $\Omega_{\vec{k}} = 2 \tilde{\mathcal{C}}^y_{\vec{k}}= -6 \tilde{\mathcal{C}}^x_{\vec{k}}$. The energy of the modes is obtained as $\xi_{\vec{k}}^2 = {\varepsilon_{\vec{k}}^2-\Omega_{\vec{k}}^2}=   16 \tilde{\mathcal{C}}^x_{\vec{0}} ( \tilde{\mathcal{C}}^x_{\vec{0}} -\tilde{\mathcal{C}}^y_{\vec{k}}) =   16 \tilde{\mathcal{C}}^x_{\vec{0}} ( \tilde{\mathcal{C}}^x_{\vec{0}} + 3\tilde{\mathcal{C}}^x_{\vec{k}})$. We have $ \tilde{\mathcal{C}}^{x}_{\vec{0}} = \sum_{\vec{r}_{ij}} \mathcal{C}_{ij}^{x} = \sum_{\vec{r}_{ij}} \frac{\Gamma \Omega^2}{12\Delta^2} \frac{\cos (k_0 |{\bf r}_{ij}|)}{\big(k_0 |{\bf r}_{ij}|\big )^3} = \sum_{j} \frac{\Gamma \Omega^2}{12\Delta^2} \frac{\cos (k_0 a j)}{\big(k_0 a j\big )^3}\to v(a)$, where $v(a=0.1\lambda) \approx 3.283$ is a constant that does not scale with $N$ at large $N$. By comparing the numerical values, we see that $\sum_{j}\mathcal{C}_{0j}^{x} \approx \mathcal{C}_{01}^{x} = 3.26 $. Then, we can approximately obtain the FTs as $ \tilde{\mathcal{C}}^{x,y}_{\vec{k}} = \sum_{j} \exp (-i k_X a j) \mathcal{C}_{0j}^{x,y} \approx \exp (-i k_X a ) \mathcal{C}_{01}^{x,y}$ for $k_X =0,\pi$. For a mode to be unstable ($\xi_{\vec{k}}^2<0$), we must have $ \tilde{\mathcal{C}}^y_{\vec{k}} > v$. For $k_X=\pi/a$, we get $ \tilde{\mathcal{C}}^{y}_{\pi/a} \approx  - \mathcal{C}_{01}^{y} = 3 \mathcal{C}_{01}^{x}>v \approx \mathcal{C}_{01}^{x} $. 

In summary, we find that due to the short-range nature of the coefficients in 1D, the instability condition is dominated by the nearest-neighbor coefficients and is always satisfied for $k_X=\pi/a$. For this to happen, it is crucial for $ \mathcal{C}_{01}^{y}<0$, i.e., the nearest-neighbor interaction coefficients along the $X$-axis of the array must be negative. This is satisfied for $\hat{e}_{\rm L} = \hat{Z}$, as shown in the first figure (schematic) of the main text. This intuition from 1D helps us understand the large growth of the corresponding mode in 2D, $(k_X,k_Z)=2\pi/a(1/2,1)$. In 2D, while the picture gets more complicated and this mode gets stabilized due to contributions from the next-nearest and the next-to-next neighbor terms, the structure of the modes is still qualitatively similar to 1D at early times, as this mode remains lowest in energy and grows the most.

\section{\textit{N}-Scaling of Spin-squeezing for 2D}

\begin{figure}[ht!]
\includegraphics[width=0.3\linewidth]{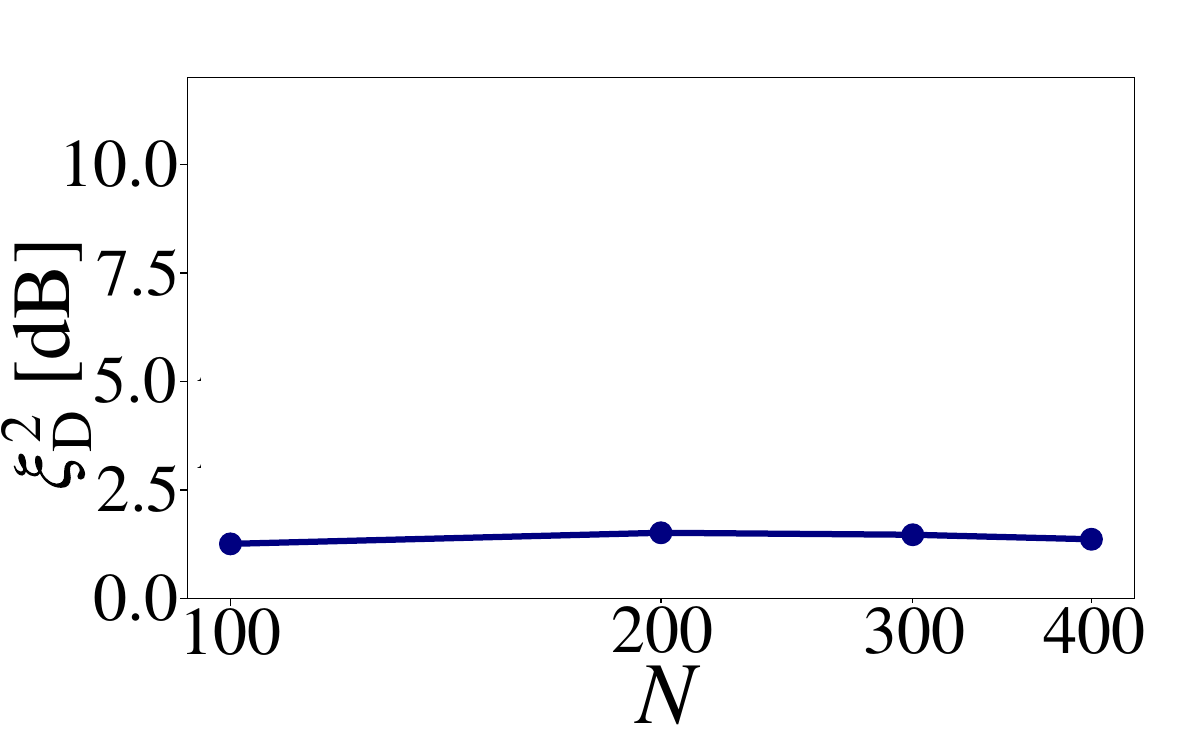} 
\includegraphics[width=0.3\linewidth]{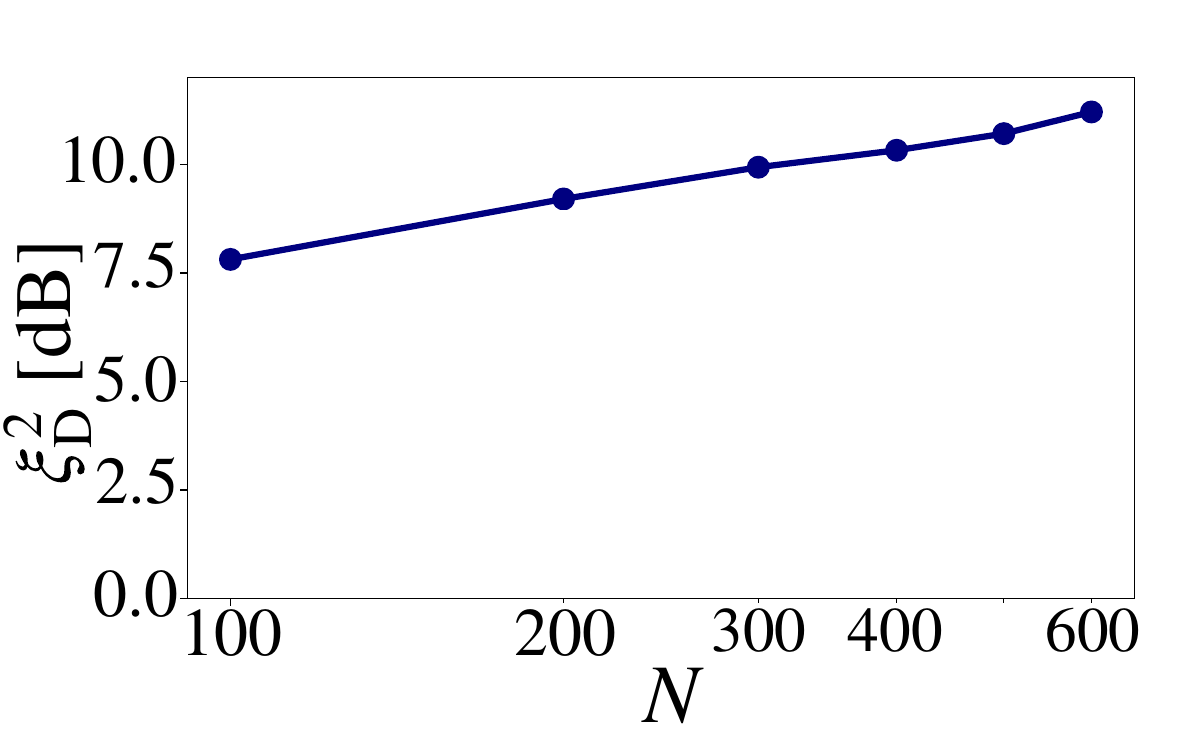} 
\caption{\label{fig:figssqueezing2Dscaling}The absolute value of the spin-squeezing parameter ($y$-axis) (Eq.~(\ref{eq:Toth_squeezing}) \cite{TothPRA2009}) minimized over time for a 2D array of $N$ atoms in the large-detuning limit using DTWA. Left: full $\hat{H}_{\rm eff}$ dynamics. Right: dynamics under a modified Hamiltonian, with the pair-creation terms turned off $(\Delta^{ij}_{1,-1}=0)$ and the interaction coefficient modified to remove the oscillating factors $(\sim \cos(k_0 r_{ij}) )$. Note that positive $y$-axis implies that the state is squeezed here.}
\end{figure}

As shown in Fig.~\ref{fig:figssqueezing2Dscaling} (left), the spin-squeezing parameter for the $\vec{k}=0$ mode (Eq.~(\ref{eq:Toth_squeezing})) does not scale with $N$ for the $\hat{H}_{\rm eff}$ in Eq.~(\ref{eq:H_eff_large_det}) for a 2D array. However, turning off the pair-creation terms $(\Delta^{ij}_{1,-1}=0)$ and modifying the interaction coefficient to remove the oscillating factors $(\sim \cos(k_0 r_{ij}) )$, does lead to a scaling with $N$ (Fig.~\ref{fig:figssqueezing2Dscaling} (right)). This shows that the growth of the collective spin-spin correlations $(\vec{k}=0)$ is hindered by the pair-creation terms and the oscillating nature of dipolar interactions.
Pair-creation reduces squeezing by leading to the growth of modes other than $\vec{k}=0$. The oscillating nature of interactions becomes relevant at larger distances, $k_0 r > \pi/2$, by suppressing the long-range correlations and has been known to prevent the growth of the $\vec{k}=0$ mode \cite{BrunoPRL2001}.

\section{Validating the Discrete Truncated Wigner Approximation (DTWA)}

\begin{figure}[ht!]
\includegraphics[width=0.35\linewidth]{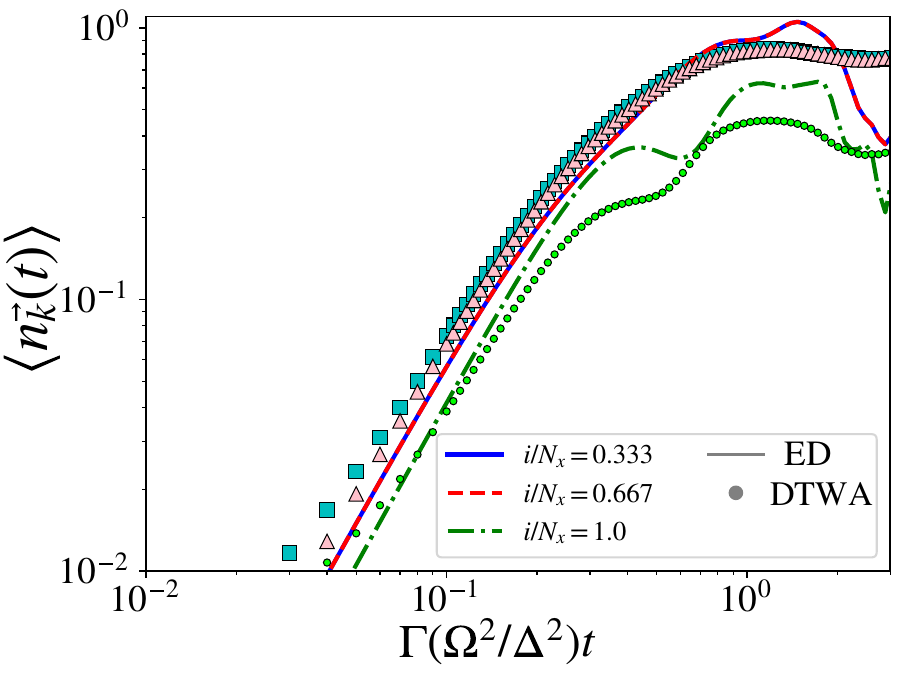} 
\caption{\label{fig:figsDTWAED} Mode occupation, $\langle n_{\vec{k}}(t)\rangle = \big(\langle \hat{\tilde{S}}_z^2 (\vec{k}) \rangle + \langle \hat{\tilde{S}}_y^2 (\vec{k})\rangle \big)/N - 1/2 $, of a $3\times 3$ array in the $X-Z$ plane with $\hat{e}_{\rm L}=\hat{e}_0 = \hat{Z}$, obtained from exact diagonalization (ED) (solid/dashed) and DTWA (dots) with fixed $k_Z=2\pi/a$ and varying $k_X=2\pi i /(N_X a)$ ($N_X=\sqrt{N}=3$).}
\end{figure}

\begin{figure}[ht!]
\includegraphics[width=0.35\linewidth]{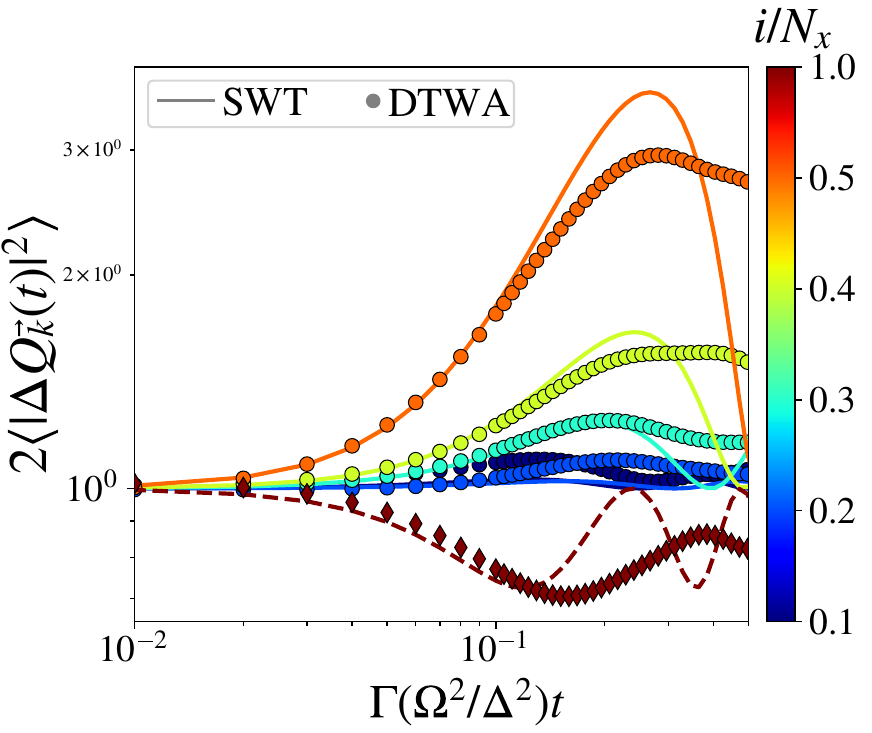} 
\caption{\label{fig:figs5} Dynamics of the mode variance for a $10\times 10$ array in the $X-Z$ plane with $\hat{e}_{\rm L}=\hat{e}_0 = \hat{Z}$, obtained from spin-wave theory (SWT) (solid) and DTWA (dots) for $k_Z=2\pi/a$ and varying $k_X=2\pi i /(N_X a)$ ($N_X=\sqrt{N}=10$). The $y$-axis is $\langle \hat{Q}_{\vec{k}}^2 \rangle (\phi = \pi/2) = \langle \hat{P}_{\vec{k}}^2\rangle= \frac{2}{N} \langle \hat{\tilde S}_y^2 ({\vec{k}}) \rangle$ .}
\end{figure}

\subsection{Comparing with ED}

In Fig.~\ref{fig:figsDTWAED}, we compare the unitary dynamics of DTWA averaged over $\sim 10^4$ trajectories with ED for a $3\times 3$ array and we find good agreement in the spin-structure factor (mode occupation) at early times. We find similar agreement for the variances of the spin-wave operators (not shown here). 

\subsection{Comparing with spin-wave dynamics}

We compare the DTWA numerics averaged over $\sim 10^4$ trajectories with the spin-wave analysis (SWA) results for a $10\times 10$ array to check the validity of DTWA. As shown in Fig.~\ref{fig:figs5}, we see good agreement of the spin variance dynamics at early times between DTWA and SWA. We find a similar trend of agreement across other observables such as the mode occupation.

\end{document}